\documentclass[twocolumn,showpacs,superscriptaddress]{revtex4}

\newcommand{\beq}{\begin{equation}}
\newcommand{\eeq}{\end{equation}}
\newcommand{\ket}[1]{\left| {#1} \right>}
\newcommand{\bra}[1]{\left< {#1} \right|}
\newcommand{\smallket}[1]{\left| \right. {#1} \left. \right>}
\newcommand{\C}{^C\!}

\usepackage{graphics}
\usepackage{amssymb}

\begin{document}

\title{Overhead and noise threshold of fault-tolerant quantum error correction}

\author{Andrew M. Steane
\\
\small Centre for Quantum Computation, Department of Atomic and Laser Physics, \\
\small Clarendon Laboratory, Parks Road, Oxford, OX1 3PU, England}

\date{\today}

\begin{abstract}
Fault tolerant quantum error correction (QEC) networks are studied by a
combination of numerical and approximate analytical treatments.
The probability of failure of the recovery operation is calculated
for a variety of CSS codes, including large block codes and
concatenated codes. Recent insights into the syndrome
extraction process, which render the whole process more efficient
and more noise-tolerant, are incorporated.
The average number of recoveries which can be
completed without failure is thus estimated as a function of
various parameters. The main parameters are the gate ($\gamma$)
and memory ($\epsilon$)
failure rates, the physical scale-up of the computer size, and
the time $t_m$ required for measurements and classical processing. The
achievable computation size is given as a surface in
parameter space. This indicates the noise threshold as well
as other information. It is found that concatenated codes
based on the $[[23,1,7]]$ Golay code give higher thresholds than
those based on the $[[7,1,3]]$ Hamming code under most
conditions. The threshold gate noise $\gamma_0$ is a function of
$\epsilon/\gamma$ and $t_m$; example values are
$\{ \epsilon/\gamma, t_m, \gamma_0 \}$ =
$\{ 1, 1, 10^{-3} \}$,
$\{ 0.01, 1, 3 \times 10^{-3} \}$,
$\{ 1, 100, 10^{-4} \}$,
$\{ 0.01, 100, 2 \times 10^{-3} \}$, assuming zero cost for information
transport. This represents an order of magnitude increase in tolerated
memory noise, compared with
previous calculations, which is made
possible by recent insights into the fault-tolerant QEC process.
\end{abstract}


\maketitle


The possibility of robust storage and manipulation of quantum information has
profound practical and theoretical implications. It would allow
highly complex quantum interference and entanglement phenomena,
including quantum computing, to be realized in the laboratory, and
it also underlies a new and as yet little understood area of
physics concerning the thermodynamics of complex entangled quantum systems.

The challenge of achieving precise manipulation of quantum
information has inspired much ingenuity, and many established
methods of experimental physics, such as adiabatic passage,
geometric phases, spin echo and their generalizations can be
useful. These provide an improvement in the precision of some
driven evolution by a given factor at a cost in speed, for example
a slow-down of the evolution by the same factor. Such methods may
play a useful role in a quantum computer, but they cannot provide
all the stability required, for two reasons. First the slow-down
is unacceptable when large quantum algorithms are contemplated,
and secondly it is doubtful whether they will in practice achieve
the relative precision of order $1/KQ$ which is needed to allow a
successful computation involving $Q$ elementary steps on $K$
logical qubits, when $KQ$ reaches values $\gg 10^6$ which are
needed for computations large enough that a quantum computer could
out-perform the best available classical computer.

Quantum error correction (QEC)
\cite{95:Shor,96:SteaneA,96:Calderbank,96:SteaneB} may allow a
precision $\ll 10^{-6}$ per logical operation to be attained in
quantum computers. In order for this to be possible, QEC must be
applied in a fault-tolerant manner, that is, the QEC process is
constructed so that it removes more noise than it generates when
it is itself imperfect. The main concepts of fault-tolerance were
introduced by Shor \cite{96:Shor}, and further insights have been
discovered by several authors
\cite{96:DiVincenzo,98:Preskill,98:GottesmanA,99:GottesmanB,97:KitaevA,98:Aharonov,98:Knill,97:SteaneA,97:SteaneC,99:SteaneB,02:SteaneA}.
Most of these studies have been concerned with the discovery of
methods which achieve fault-tolerance in a quantum computer, and
with finding scaling laws which describe how the tolerated noise
level varies with the length of the computation. In this paper I
address the problem of estimating the amount of noise which can be
tolerated, and quantifying the cost of the stability in terms of
the required increase in the number of physical qubits in the
computer.

Some previous efforts to answer these questions have concentrated
on the idea of the {\em threshold}. This is the result that
arbitrarily long quantum computations can be robust, under various
reasonable assumptions, once the noise per quantum gate and per
qubit during the duration of a gate is below a threshold value
which does not depend on the size $KQ$ of the computation
\cite{98:Knill,98:Aharonov,97:KitaevA,97:KitaevB,Th:Gottesman,98:Preskill}.
Estimates of the value of this threshold have varied between
$10^{-6}$ and $10^{-3}$, in the case that gates can act between
any pair of qubits in the computer. In view of this wide range, a
new calculation of the threshold is valuable, and is one of the
aims of this paper. The discussion will include various issues
such as the speed of measurements and classical processing, and
the best choice of encoding, which have been not been addressed up
till now.

However, the threshold result is of limited practical
significance, because the encoding it requires (namely multiply
concatenated coding) fails to take advantage of a fundamental
property of error correction theory, which is the existence of
{\em good} codes. These have rate $k/n$ and relative distance
$d/n$ both bounded from below as the block size $n$ increases;
they allow error-free transmission of
information at a rate close to the channel capacity. Once the
noise is brought moderately below the highest threshold
offered by multiply concatenated codes,
good encodings (which do not have a threshold result) allow
very large quantum algorithms to be stabilized at a much lower
cost in scale-up of the physical resources (qubits and operations)
of the computer. The only existing estimate \cite{99:SteaneB} of
what these good codes can achieve used a simple analysis which is
only valid in the limit of low noise rate, and which does not take
advantage of recent insights into the syndrome extraction process
\cite{02:SteaneA}. It remains difficult to compare this estimate
with the threshold calculations, because each depends on the noise
model and the way the noise rate is parameterized, and different
authors make different choices. The present paper treats both
unconcatenated and concatenated codes together, and so permits a
comparison between them.

A central concept which emerges from this uniform treatment is
to regard the maximum
computation size $KQ$ which can be stabilized
as itself a function of various parameters. These parameters divide into
two types.  The first type quantifies the noise and imprecision
which can be tolerated, the second type quantifies the demands
on the physical hardware, such as the degree of parallelism
and especially the redundancy or scale-up (increase in number of qubits)
required. Hence $KQ$ is best understood as a surface, i.e. a function
of two main parameters: the tolerated noise level and the physical
scale-up. The threshold result is an interesting asymptotic
behaviour of this surface in the region of high noise and scale-up, but
what we would like to know, and what is also here discussed, is
the form of the surface elsewhere in parameter space.

These questions are here addressed by numerical simulations of
quantum error correcting networks, and by a detailed approximate
analysis.

The paper is laid out as follows. The basic concepts of
fault-tolerant quantum computing are briefly sketched in section
1, and the noise model adopted in the rest of the paper is
described. Section 2 gives the complete protocol for QEC,
explaining various choices about the way the networks are
constructed. Section 3 presents the results of numerical
simulations of these networks for the case of the $[[7,1,3]]$
Hamming code and the $[[23,1,7]]$ Golay code. Section 4 gives an
analysis of the noise and error propagation in the QEC protocol.
The numerical results are used to provide values of two fitted
parameters, and to confirm the correctness of the general trends
predicted by the analysis. The results of the analysis are then
presented for 18 different quantum codes, correcting between 1 and
9 errors, and encoding between 1 and 43 qubits per block.
Section 5 adapts the analysis to the case of
concatenated coding. The results of concatenating once are
presented, and the threshold associated with multiple
concatenations is calculated. Section 6 then describes and
discusses the $KQ$ surface.

Fault-tolerant computation and fault-tolerant data storage are
largely similar in that the recovery operations dominate the
dynamics. Nevertheless there is a distinction between them. The
present treatment is thorough for the case of data storage, and it
is argued in section 3 that a judicious placement of logic gates
in between recoveries allows the case of fault-tolerant
computation to be like data storage with simply some additional
noise from those gates. Therefore the present results apply to
computation (not just data storage). However a more thorough
treatment of the error propagation directly between data blocks is
needed in order to clarify this point.

The main results are as follows. First, the threshold for quantum
computing using multiply concatenated coding is higher when the
code is based on the $[[23,1,7]]$ quantum Golay code rather than
the $[[7,1,3]]$ Hamming code. The former also requires a lower
scale-up at given $KQ$ than the latter so is advantageous for both
reasons. It is found that the time taken to complete measurements
and classical processing on qubits is also a significant factor
which has mostly been overlooked in previous treatments. When the
noise $\epsilon$ per memory qubit per gate time  is the same as
the noise $\gamma$ associated with a gate, and the measurement of
a qubit takes the same time as a quantum gate, the threshold is
$\gamma_0 = \epsilon_0 \simeq 10^{-3}$. If the measurement takes
100 times longer than a gate, the threshold is $\gamma_0 =
\epsilon_0 \simeq 10^{-4}$. When the noise per memory qubit per
gate time is 100 times smaller than that of a quantum gate, the
threshold is $\gamma_0 = 100 \epsilon_0 \simeq 2 \times 10^{-3}$
(see figure 7 for more information).

The complete $KQ$ surface, plotted on logarithmic axes, is found
to have the shape approximately of a set of inclined planes
separated by steep cliffs, revealing quasi-threshold behaviour in
scale-up as well as noise (figures 8--10). The jumps in $KQ$ as a
function of scale-up occur when new types of encoding become
possible. When the noise is an order of magnitude below threshold,
and memory is much less noisy than gates,
a scale-up of order 10 permits $KQ$ up to $\sim 10^{10}$ by using
good codes such as BCH codes.
At a scale-up of order 1000, $KQ$ up
to $\sim 10^{40}$ is available by using a good code concatenated
once with the $[[23,1,7]]$ Golay code.
If the memory is as noisy as the
gate operations (which could be the case, for example, when
information is moved around using swaps between neighbouring
bits), a larger scale-up or smaller gate noise is required.

\section{Basic concepts}

A quantum computer stabilized by QEC methods has three stages in
its operation. First there is a preparation stage, which places
the computer in a close approximation to the state
$\left| \right.{ 0^{(K)}} \left. \right>_L$ which is
the logical zero state of the $K$
logical qubits of the computer. Then there is a sequence of $Q$
logical operations, interspersed with error correction (also
called {\em recovery}) of the whole computer. Then the
individual physical bits of the computer are measured in the
computational basis, and a final error-correction is applied by
classical computation to the classical data thus acquired. The
overall probability of success is the probability that the
classical bit string obtained at the end of this final recovery
represents a correct solution to the computational problem being
addressed.

For the initial preparation stage, a sufficient approximation to
$\left| \right.{ 0^{(K)}} \left. \right>_L$
can be obtained by a fault-tolerant
measurement of the logical state of all the logical qubits,
combined with an error correction \cite{99:SteaneB}, followed by
fault-tolerant gates to flip logical bits which were found to be
in the logical 1 state.

The final classical correction can be represented in an abstract
way as an operation $\cal R$ on the density matrix $\rho(Q)$ of
the computer after the $Q$'th computational step. Then a suitable
measure of success of stabilization by QEC is the fidelity $F_Q
\equiv \bra{\psi(Q)} {\cal R} (\rho(Q)) \ket{\psi(Q)}$, where
$\ket{\psi(Q)} = U_Q \ket{0^{(K)}}_L$ is the ideal (i.e. noise and
imperfection-free) state of the computer after a sequence of $Q$
perfectly-executed elementary steps.

An exact calculation of $F_Q$ is extremely difficult, and cannot
be attained for a system of even just a few
logical bits and operations, owing to the complexities of the
encoded states and of the interactions of the physical qubits with
each other and the rest of the world. In this paper
$F_Q$ will be estimated by adopting a very simple noise model and
performing a numerical and combinatorial analysis of the QEC
networks.

The computer will be encoded using a quantum error correcting code
of parameters $[[n,k,d]]$ where $n$ is the number of physical
qubits per block, $k$ is the number of logical qubits per block
(which N.B. can be greater than 1) and $d$ is the minimum distance
of the code. The code is $t$-error correcting where $t=(d-1)/2$.
The networks to perform recovery will be built
according the recipe put forward in \cite{99:SteaneB,02:SteaneA,02:SteaneB}, which
I will outline in section \ref{s:protocol}.

\subsection{Noise model}  \label{s:noise}

`Noise' in the context of QEC is taken to mean any process which causes
the state of the physical qubits of the computer to be different
from what it should ideally be \cite{96:SteaneB,01:SteaneA,97:Knill,00:KnillB}. Thus
we include undesired
interactions between the qubits, and terms in their internal
Hamiltonian and in their coupling to the environment
which are known to be present but which cause undesired effects,
as well as further terms whose details may be unknown us, all
under the umbrella concept of `noise'. It is an established
feature of QEC that the overall effect of noise can be understood
in terms of the set of Pauli operators and the identity acting on
the physical qubits. I will write these operators $I,X,Z$ and
$Y=XZ$. It is convenient to define the $Y$ operator
so that it is real, it then differs from the Pauli operator
$\sigma_y$ by a factor of $i$ which does not affect the argument.

It is important to distinguish between the processes which cause
imperfection in the computer state, which I will call `failures',
and the resulting imperfections in the state, which I will call
`errors'. For example, a single failure of a two-qubit gate can
result in two errors, meaning the state after the failure involves
errors in two of the physical qubits (that is, a tensor product of
Pauli operators on both qubits is required to restore the state). In
general after the action
of some quantum network, a single failure somewhere in the network
can result in multiple errors. The main feature of
`fault-tolerant' networks is that a single failure anywhere in the
network leads to only one error (or an acceptable number of
errors) per encoded block. A set of $m$ single-bit errors
on $m$ qubits will also
be referred to as an error of {\em weight} $m$.

When the noise produces an effect large enough that the computer
state cannot be corrected by QEC, the whole quantum computation
must be assumed to fail, since it is close to
certain that it will not produce a useful result ($F_Q \simeq 0$). This
situation will be called a {\em crash}. QEC and fault-tolerant gate
methods allow the crash probability to be much smaller than
the failure probability of individual elementary operations on the
physical qubits.

The noise model which I will adopt for the purpose of estimating
$F_Q$ is as follows. At each time step, every freely evolving
physical qubit has no change in its state with probability
$1-\epsilon$, or undergoes rotation by the operator $X,Y$ or $Z$
with equal probabilities $\epsilon/3$. Such failures are termed
`memory failures' and $\epsilon$ is the memory failure
probability. Every gate is modeled by a failure followed by
a perfect operation of the gate. The failure for a single-qubit
gate is the same as a memory failure except that it occurs with
probability $\gamma_1$. The failure of a two-qubit gate is
modeled as a process where with
probability $1-\gamma_2$ no change takes place before the gate, and with equal
probabilities $\gamma_2/15$ one of the 15 possible single- or
two-qubit failures takes place (these are
$IX,IY,IZ$, $XI,XX,XY,XZ$, $YI,YX,YY,YZ$, $ZI,ZX,ZY,ZZ$).

Every preparation of a single physical bit in $\ket{0}$ will be
modeled as a perfect preparation follow by a single-bit failure
of probability $\gamma_p$. Every measurement of a single physical
qubit will be modeled as a single-qubit failure of probability
$\gamma_m$, followed by a perfect measurement. Such a model
accounts satisfactorily for the main ways in which measurements can fail,
with this exception: a qubit measurement might give a certain eigenvalue $\lambda$
as measured outcome, but the qubit is not projected into the corresponding eigenstate
$\ket{\lambda}$. In
the present context, however, the latter case is equivalent in its
effects to the case which is modeled (i.e. failure followed by
perfect measurement), because the measurements are always used to
acquire syndrome information. All that matters is that the
measured eigenvalue either does or does not correctly indicate the
error in the computer: this is accounted for by the model.
The case where the syndrome bit was projected
onto a state other than $\ket{\lambda}$ does not have any
further impact on the computer because we never re-use measured
bits without re-preparing them in $\ket{0}$ (a process which has its own
failure probability $\gamma_p$).

`Leakage' failures, which occur when the physical computer moves
out of the Hilbert space spanned by the physical qubits, are
assumed to be suppressed by techniques such as optical pumping
or small leakage measuring networks \cite{98:Preskill}
and hence converted into failures of the type already considered.
The leakage probability is absorbed into the gate and memory failure
probabilities.

The model is defined so that qubits participating in a gate in a
given time-step undergo gate noise but not memory noise. In other
words, the gate noise parameters $\gamma_i$ are defined in such a
way that they include all the noise acting on the qubits
participating in the gate during the time of action of the gate.
It is necessary to be explicit about this distinction for the
calculation of thresholds in section \ref{s:thresh}.

The QEC networks I will analyze are composed only of
the single-qubit Hadamard transform and two-qubit controlled-not or
controlled-phase gates, and state preparation and measurement of single
qubits in
the computational basis.

An implicit assumption of this noise model is that failures are
uncorrelated and stochastic. The first assumption
(uncorrelated failures) can be relaxed
without significantly changing the overall results as
long as correlated failures have probability sufficiently smaller
than uncorrelated ones. In a single time-step, uncorrelated memory
failures in $n$ qubits
give $m$-bit errors with probability
\beq
B(n,m,\epsilon) \equiv \frac{n!}{m! (n-m)!} \epsilon^m
(1-\epsilon)^{n-m}.   \label{B}
\end{equation}
If correlated failures (for example due
to $m$-body interactions between the physical qubits) have a
probability small compared to this then they can be neglected
in a calculation of the crash probability without
significantly affecting the result.
Unwanted systematic effects in a computing device will also cause a
finite correlation between the failures in nominally independent
gate operations, but if the probability for a weight-$m$ error to be
produced by correlated gate failure is small compared to the probability
that the same error is produced by uncorrelated gate failures, then it is
sufficient to analyze the latter.

Similar statements can be made about non-stochastic contributions
to the noise. An example is rotation errors: if a given qubit
is erroneously rotated $m$ times by a small angle $\theta$,
then if the angles are all in the same direction they add
coherently to give a net angle $m \theta$ and error probability
$\sim m^2 \theta^2$, whereas if the direction of rotation is random, a
random walk is produced resulting in a mean net rotation $\sqrt{m}
\theta$ and overall error probability $\sim m \theta^2$. The model
treated here assumes the latter case; this will cover the main
features as long as the coherent contribution gives a net error
similar to or less than the incoherent one for each application
of the recovery network.

Recently Alicki {\em et al.} \cite{01:Alicki} have drawn attention
to another implicit assumption, namely that the noise is
independent of the dynamics of the recovery network, which they
show is false for quantum reservoirs with long-range `memory' (such
as electromagnetic vacuum). This implies the noise is both
correlated and non-stochastic. The argument is
subtle and it remains an open question whether the
structure of the correlations is of a type which defeats fault-tolerant
QEC, or has an influence small compared to the stochastic
uncorrelated part which I will
estimate here.

\section{Correction protocol} \label{s:protocol}

A fault-tolerant error correction can be accomplished with a
variety of choices of exactly how the syndrome extraction network
is constructed. Here I will make choices which I have previously
argued to be close to optimal, when considerations of noise
tolerance and the overall required scale-up are both taken into account.

{\em Transversal} operation of a gate means the gate is applied
once to each physical qubit in a block, or once to each
corresponding pair of physical qubits in a pair of blocks for the
case of a 2-qubit gate. {\em Blockwise} action of an operator
means the operator is applied once to each logical qubit in a
block, or once to each corresponding pair of logical qubits in a
pair of blocks for the case of a 2-qubit gate.

The QEC code will be a CSS code obtained from a classical code
which contains its dual. Such codes have the property that
transversal controlled-not and controlled-phase operations act as
blockwise controlled-not and controlled-phase operations
respectively, and transversal Hadamard acts as blockwise Hadamard
\cite{98:GottesmanA,99:SteaneB}.
A further property is useful for constructing fault-tolerant
logical operations, though it is not needed for fault-tolerant
QEC. This is the property that the underlying classical code is
doubly even (i.e. the codewords have weights a multiple of 4)
\cite{96:Shor,98:GottesmanA,99:SteaneB}. I
will restrict attention here to such codes.

If the algorithm to be accomplished requires $K_I$ qubits on an
ideal (noise-free) machine, then the real computer has $K$ logical
qubits encoded in $K/k$ blocks, each block consisting of $n$
physical qubits, where $K$ is larger than $K_I$ by a fixed small
amount which can be $< 10$. The few extra blocks are necessary
as workspace to allow fault-tolerant logical operations on the
logical qubits using methods such as teleportation.

For each such `data block' the computer contains in addition $2
n_{\rm rep}$ ancilla blocks of $n$ physical qubits each, and $2
n_{\rm rep}$ sets of verification bits, each set containing
$(n+k)/2$ physical qubits.
The total number of physical qubits in the computer is thus
   \beq N = \left( n + n_{\rm rep} (3 n + k) \right) K/k.
   \label{N}
   \end{equation}
$n_{\rm rep}$ is the number of pairs of
ancilla blocks per data block which can be prepared in parallel, in order to
speed up syndrome extraction; it will have a value typically in
the range 1 to 10.

The verification bits are used to verify prepared ancilla states.
The stabilizer of the zero state $\left| \right. {0^{(k)}} \left.
\right>_L$ of a single block (i.e. $k$ logical bits) is generated
by a set of $n$ linearly independent operators. This set can be
expressed such that it divides into a subset of $(n-k)/2$ which
consist of tensor products of $X$ operators, and $(n+k)/2$ which
consist of tensor products of $Z$ operators. The ancilla state is
verified once against $X$ errors only, by measuring the
eigenvalues of the latter subset (the one composed of $Z$
operators) using the verification bits. It is proved in
\cite{02:SteaneA} that this single verification is sufficient
to produce the correct fault-tolerant behaviour when the detailed
form of the set of stabilizers is properly chosen.

A single complete recovery consists of $X$-error correction and
$Z$-error correction. These two halves of the correction proceed
in parallel. While the $X$-error correction machinery is preparing
ancilla states, the $Z$-error correction machinery is coupling its
ancillas to the data blocks, and vice versa. A single complete
$X$-error correction of a single data block proceeds as follows, and
the $Z$-error correction is identical except where
indicated (for diagrams see \cite{97:SteaneC,99:SteaneB,02:SteaneB}).
Correction of different data blocks proceeds in parallel.

\begin{enumerate}
\item Prepare $n_{\rm rep}$ ancilla blocks in $\smallket{0^{(n)}}$.
\item Operate a network $G$ in parallel on each of these ancilla
blocks. $G$ consists of Hadamards and controlled-not gates, and
if perfect would produce the transformation $\smallket{0^{(n)}}
\rightarrow \smallket{0^{(k)}}_L$.
\item Using verification bits prepared in $(\ket{0} + \ket{1})/\sqrt{2}$,
verify the ancilla blocks by operating a network $V$ consisting
of controlled-phase gates between each ancilla block and its verification
bits, followed by Hadamard transformation of the verification
bits and their measurement in the computational basis.
\item Ancilla blocks which pass the verification (i.e. all
verification bits were found in the state $\ket{0}$ when measured)
are deemed `good' and are used in the rest of the protocol.
Those that do not are left alone until they are re-prepared
at the beginning of the next round of QEC. Let $\alpha$ be the
fraction which are good, so that we now have $\alpha n_{\rm rep}$
good ancillas.
\item Couple 1 `good' ancilla to the data block by
blockwise controlled-phase (for $X$-error correction)
or controlled-not (for $Z$-error correction), with the ancilla
acting as control, the data as target.
Hadamard transform this ancilla block and then measure each of
its physical qubits in the computational basis.
Use a classical computer to decode the classical
bit string thus obtained, and hence derive the error syndrome
\cite{97:SteaneA,97:SteaneC}.
\item If this syndrome is zero, no further action is taken. The
data block rests until recovery has been completed on all the
data blocks in the computer whose first syndrome was not zero.
Let $\beta$ be the fraction of blocks which give a zero syndrome.
\item (a) Otherwise, couple $r-1$ further good ancillas to the data
block by blockwise
controlled-phase (controlled-not), for $X$-correction ($Z$-correction),
where $r$ is a parameter to be optimized. Hadamard
transform and measure these ancillas in parallel, as in step 5.
\item (a) We now have a total of $r$ syndromes extracted for each data
block whose first syndrome was non-zero. We accept any group of $r'$
syndromes which all agree, where $r'$ is a parameter to be
optimized. When a syndrome
is accepted, the data block is corrected accordingly by
application of one or more $X$ gates (or $Z$ gates). If no
acceptable syndrome is found, no further action is taken, so the
data block goes uncorrected for $X$ errors ($Z$ errors)
in this round of QEC.
\end{enumerate}

Steps 7(a) and 8(a) will be modified below, but to understand
the modification it helps to begin with the statements as given.

The syndrome repetition factors $r$ and $r' \le r$ will be chosen so
as to maximize the probability of success. Increasing $r'$ reduces the
probability of accepting a wrong syndrome, but increasing $r$
increases the noise accumulating in the data block.
The $\alpha n_{\rm rep} - 1$ good ancillas per data block
which were not used in step 5 are sufficient to allow
$r \le r_{\rm max} = 1 + (\alpha n_{\rm rep} - 1) / (1-\beta).$

In the protocol described we have $\alpha n_{\rm rep}$ good ancillas
per data block during each round of QEC, and we require on average
$\beta + r (1-\beta)$ for one correction. Hence we have enough
ancillas to complete $\alpha n_{\rm rep}/({ \beta + r (1-\beta) })$
independent corrections almost in parallel\footnote{They cannot be
completely in parallel because the data block can only be coupled
to one ancilla block at a time. This is the most reasonable
assumption, because it must be arranged that successive syndromes
have independent noise, so it is not sensible to try
to couple one data block to many ancillas by a single operation.}.
The sequential part is
the gates which couple data and ancilla. Increasing $n_{\rm rep}$
reduces the time during which the data is left alone between
corrections, so is valuable when the memory noise accumulating directly
in the data contributes a significant part of the total data errors.
However, much of the error in the data arises by propagation
from the ancillas, or from the gates coupling data and ancillas,
and these contributions are unaffected by $n_{\rm rep}$.

We will mostly be interested in the case of large quantum
algorithms, for which the failure rates must be small so $\alpha$
and $\beta$ are close to 1, and the number of corrections in
parallel is close to $n_{\rm rep}$. The exception is when a
concatenated code is being used, with error rates close to the
threshold. In this case $\alpha$ and $\beta$ can be of order $0.5$
for the innermost levels of the concatenated coding hierarchy,
therefore $n_{\rm rep}$ must be increased to allow sufficient
ancillas for rapid correction.

The protocol can be refined primarily in two ways. First, one can
operate a different and possibly more sophisticated scheme to
prepare and verify ancillas in step 3, and secondly one can
adopt a more sophisticated response to the syndrome information in
steps 7 and 8.

For example, in step 3 one could verify the ancilla twice
and accept if it passed at least once, or one could prepare two
ancillas and then
compare them by a transversal controlled-not followed by measurement of
one. The former case requires more time, which can be compensated
by an increase in $n_{\rm rep}$, and the second case requires more
ancillas. However, any attempt to improve the ancilla preparation
can only result in a modest reduction of the crash probability (at
given noise rate) because the gates connecting ancilla and data
cause much of the noise in the data, and these cannot be
avoided. This is discussed after equations (\ref{gr}), (\ref{sr}) below.

An example of a more sophisticated procedure in step (8) is to
extract more syndromes immediately if insufficient syndromes agree.
In his calculation, Zalka \cite{99:Zalka} employed refinements of
this kind. However, such a response is only valuable if it can be
made quickly, and this requires fast measurements. It is
physically reasonable to suppose that measurement of a qubit may be slow
compared to one time step, where a time step
is the duration of a two-qubit gate. When measurement is slow
it is better
to couple syndrome information into ancillas as many times as will be
required all at once, and then measure the ancillas in parallel.
Therefore if one wishes to extract one further syndrome in step (8)
when insufficient syndromes agree, it is advantageous to
extract further syndromes as well and one has in the end a protocol
close to the one being considered.

There is a modification to steps 7 and 8 which is worth making since
it requires only a slight change in the classical part of the processing
so has negligible cost. This is to improve the case where no acceptable
syndrome was found for a given block. In this case, at the next
recovery rather than extracting a further $r$ syndromes, we extract
$r'' \le r$ where $r''$ is another parameter to be optimized, and then
make the best use of the $r+r''$ syndromes
available from the most recent extractions. Typical values for
$r''$ are in the range $r/2$ to $r$.

\begin{enumerate}
\setcounter{enumi}{6}
\item (b) In the case that at the last recovery, sufficient
syndromes were found in agreement for the block to be corrected
for the error-type under consideration, proceed as in
7(a). Otherwise, now extract $r''-1$ syndromes.
\item (b) In the case that at the last recovery, sufficient
syndromes were found in agreement for the block to be corrected
for the error-type under consideration, proceed as in
8(a). Otherwise, now examine the $r+r''$ most recent syndromes
obtained from this and previous recovery attempts. Accept any group of
$r'$ syndromes which all agree, giving preference
to more recently extracted syndromes if there is more than one
acceptable group. If there is an acceptable set of syndromes,
correct the data block accordingly, otherwise do nothing.
\end{enumerate}

This reduces the noise in the data by making better use of
the syndrome information. Further refinements are possible, for
example to adjust the case where three successive extractions were
necessary, but in any case this is already a small adjustment so
there is not much further improvement available.

\subsection{Number of recoveries per computational step}
\label{s:number}

It might be thought that when the recovery time $t_R \gg 1$, which is
typically the case, it
would be advantageous to allow many logical gates to operate per
recovery, as was argued by Zalka \cite{99:Zalka}.
However, if the logical gates are not on
independent bits, then it is dangerous to allow many of them
between recoveries or the error propagation will start to avalanche.
Also, it might be argued that sometimes it is only necessary to
recover some of the blocks. However, typically the recovery time
is long enough that noise accumulating in all blocks is such
that they all need correcting.
Therefore the choice adopted here is that the whole computer must be
recovered after any simple logical gate such as controlled-not
or Hadamard is applied. On those occasions in
a given algorithm
where many logical gates can act simultaneously, then they are
implemented in parallel, followed by one complete recovery.

The logical gates are accomplished in a fault-tolerant manner
by sequences of appropriately chosen gates and measurements
\cite{98:GottesmanA,99:SteaneB,99:GottesmanB}. To quantify the algorithm size $KQ$ precisely, we
must be specific about what type of gate we are counting, because
some are easier to accomplish than others. For example, a
fault-tolerant network for a Toffoli gate may require 8
recoveries, while a controlled-not gate may only require 1 or 2.
Since the main quantity to be calculated is the crash probability
per recovery of a single block, the ``computation size" will be
taken to be the number of such recoveries when a code with
$k=1$ (one logical bit per block) is used. Codes with $k >1$
require more recoveries because the fault-tolerant constructions
are slightly more complicated. It can be shown that for standard
logical gates such as controlled-not and Toffoli, networks
for $k > 1$ exist which involve approximately twice as many
recoveries as similar networks for $k=1$, therefore to
make a fair comparison it will be assumed here
that for a given algorithm, on average
twice as many recoveries are needed when $k>1$ than when
$k=1$.

\subsection{Timing and non-nearest-neighbour coupling}

The correction protocol involves networks $G$ and $V$ for
preparing and verifying ancillas, measurement of sets of bits, and
transversal controlled-gates between ancilla and data blocks. The
precise set of operations in $G$ and $V$ is mostly dictated by the
structure of the code, with some moderate room for flexibility
in the time ordering of gates and in which set of linearly
independent parity checks is
chosen. The total time taken by the operations, by contrast,
and hence the memory noise,
is dictated not only by the logic of the network, but also by the
physical capabilities of the computing device. It will be assumed
here that {\em the computing device is capable of all the parallelism
which is logically available in the QEC protocol}. Parallel
operation of two or more gates is logically available when the
gates commute, so that their effect is the same when they are
applied all at once or sequentially.
For example, the assumption implies that a transversal gate
operation takes a single time-step, and that
parallel operation is physically available for
sets of gates within the $G$ and $V$ networks, which is useful
for speeding up the ancilla preparation.

The $G$ and $V$ networks are related to the generator matrix
and parity check matrix of the classical code $C$ whose codewords $u$ give
the state
  \beq
\ket{0^{(k)}}_L = \sum_{u \in C} \ket{u}.  \label{C}
\end{equation}
Let $H$ be
the check matrix of $C$, then the parallelism available in
the $G$ and $V$ networks was shown in
\cite{02:SteaneA} to allow the controlled-gates in
these networks to be completed in $w$ and
$w+1$ time steps respectively, where $w$ is
the maximum weight of a column or row of the matrix $A$ given by
$H = (I \, A)$ where $I$ is the $(n+k)/2 \times (n+k)/2$ identity
matrix. A further time interval is required for the Hadamard
operations and single-bit measurements and state preparations.

Consider the case that 2-bit gates such as controlled-not are only
available in the physical computer between neighbouring physical
bits. In this case we have to allow some time, and associated
noise, for the transport of the physical qubit information from
one place to another in the computer. A reasonable rough model of this
is to suppose that the speed and precision of a gate between qubits
initially separated by distance $s$ scales as $1 + s/D$, where the 1
accounts for the cost of the nearest-neighbour gate, and $s/D$
accounts for the cost of bringing the bits together from distance
$s$. In this model, $D \simeq 1$ is a reasonable estimate for a computer
which transports information by repeated swap operations between fixed
physical qubits, and $D \gg 1$ describes a
computer which can move information around at little cost.
In the QEC network, physical gates are mostly between qubits which can be
fairly close together, such as within part of one block, so a
value $D \sim 100$ is sufficient to allow $D$ to be large compared
to the mean distance $\bar{s}$ spanned by 2-qubit gates involved
in the QEC network \cite{02:SteaneB}. In the estimates to
follow, I will make the simplifying assumption of ignoring the cost of the
physical separation between physical qubits. The results for the noise
tolerance will therefore be valid only when $\bar{s}
\ll D$. I can use the results to roughly estimate what will happen for a
computer having smaller $D$ by dividing the tolerated error rates
by $1 + \bar{s} / D$. Calculations of $\bar{s}$ for two
quantum error correcting codes are described in
\cite{02:SteaneB}.

Another timing consideration is involved in the measurements and
the classical processing of the syndromes. It is an important
assumption that the verification bits and the error syndromes are
in fact measured, and not treated by purely unitary networks. This
allows a substantial part of the processing of this information to
be done classically, which I assume is both fast and precise. The
time involved in measuring a physical qubit and completing
classical processing on the measured eigenvalue will be assumed to be
$t_m$ time-steps, where 1 time-step is the time required for a
controlled-not (or controlled-phase) operation. Typical values for
$t_m$ are in the range $1 \le t_m \le 100$, which may be
associated mostly with the measurement time, making the assumption
that the classical processor has a clock rate much faster than
that of the quantum processor.

\section{Numerical calculations}  \label{s:num}

The effects of noise and error
propagation in the protocol described in section \ref{s:protocol}
were numerically calculated.
It is possible to do this in an efficient way because it is
sufficient to keep track of the propagation of the errors, rather
than the evolution of the complete computer state.

The C++ program keeps an array of $2 n + (n+k)/2$
binary digits representing $X$ errors in the physical qubits of
one data block
and one ancilla with its verification bits, and a
similar array representing $Z$ errors. Failures are generated
randomly in every gate and time-step, according to the model
described in section \ref{s:noise}, by adding 1 to members of
the $X$ and/or $Z$ error arrays at the locations of those qubits
experiencing an $X$ and/or $Z$ failure respectively. The ancilla
bits are re-used for the repeated ancilla preparations and for the
$X$ and $Z$ syndrome extraction, but memory noise is added to the
data block appropriate to the amount of time passing when
$2 n_{\rm rep}$ ancillas are available in parallel.

The action of each quantum gate in the networks is modeled by
first producing random failure, using the model described
in section \ref{s:noise}, and then accounting for error
propagation. The error propagation part is as follows:
a Hadamard gate on a single qubit
swaps the $X$ and $Z$ error values for that qubit;
a controlled-not gate adds the $X$ error of
the control bit to the target bit, and the $Z$ error of the
target bit to the control bit; a controlled-phase gate
adds the $X$ error value of the control bit to the $Z$
error value of the target bit, and the $X$ error value
of the target bit to the $Z$ error value of the control bit.

It was found that a good pseudo-random number generator was needed
in order to get reliable results at low crash probability.
For example the generator `ran0' in \cite{Bk:Press} was
inadequate; `ran3' was used instead.

The network of gates is obtained directly from the check and
generator matrices of the relevant classical codes, see appendix
for details. The gate failures were added at the locations in space and time
of the relevant gates. The memory failures were not modeled exactly
in the right way however. To save program time, during the $G$
and $V$ networks, rather than adding memory failures only to
those bits not involved in a
gate at a given time, memory failures were
distributed randomly amongst all the bits, with probabilities
set so that the mean number of failures was correct.
This change is not likely to affect the precision of the
final result, which in any case can only be compared to physical
examples in an approximate way owing to the simple noise model.

The noise caused by logical operations on the data was partially
modeled by adding a further gate failure to each qubit in the data
between each round of QEC. This completely accounts for
single-block gates but not the error propagation between data
blocks caused by logic gates between data blocks. However,
at any stage typically only a few logical qubits are involved
in 2-bit gates, and these can be timed so as to keep error
propagation to a minimum, as follows.
If a controlled-phase logic gate is to be implemented,
it should be placed just after $X$ error correction
on both blocks involved, since at this stage in their evolution
the blocks temporarily have a minimal number of $X$ errors,
and only this type of error is propagated (into $Z$ on the other
block) by the gate. Similarly, if a controlled-not gate is to be
implemented, it should be placed just after the
control bits have had $X$ error correction, and the target bits
$Z$ error correction. The cost of this is that sometimes
one or a few blocks have to
wait a little longer before being corrected, so that memory noise
occurring directly in the data block accumulates for longer.
However, since this noise is not the main source of data errors, the
omission of this detail from the numerical simulation
is not expected to affect the final result significantly.

A single logical step consists of a single transversal gate acting
on the data block, followed by the complete QEC protocol. The
program repeats this $Q$ times to represent an algorithm of $Q$
steps. After each step, the $X$ and $Z$ bit-error arrays are
examined to see if the accumulated noise represents an
uncorrectable error. If an uncorrectable error has occurred, the
run is stopped and a record is kept of how many steps were
completed successfully. This is repeated a large number (millions)
of times and the relative frequencies of success or a crash are
used to obtain estimates of the fidelity of a quantum computer
stabilized by QEC, see below. This is a type of Monte-Carlo
simulation.

The numerical calculations were carried out for two example codes,
the $[[7,1,3]]$ single-error correcting code obtained from a
classical Hamming code, and the $[[23,1,7]]$ three-error
correcting code obtained from a classical Golay code. The
classical codes in both these examples are perfect, so their quantum
versions perform especially well.

In order to interpret the $X$ and $Z$ bit-error arrays to discover
whether they represent an uncorrectable error, it is necessary to
recall the properties of quantum codes. The combination of the $X$
and $Z$ errors represents an error operator $E$ which has acted on
the data qubits. However, the weight of $E$ does not in itself
determine whether $E$ is correctable. For example, if $E$ is in
the stabilizer then it constitutes no error at all. It is necessary
to determine rather whether $E \ket{\psi}_L$ would be decoded to
$\ket{\psi}_L$ by a perfect recovery of the computer. To do
this I calculate the
syndromes $H E_X$ and $H E_Z$ where $E_X$ and $E_Z$ are the
bit-strings representing the $X$ and $Z$ parts of $E$
respectively, and $H$ is the parity check matrix of the classical
code $C$ (equation (\ref{C})). Each syndrome has a coset leader,
which is the minimal weight error vector which can cause that
syndrome. If the weight of the coset leader for either syndrome is
greater than the number of errors correctable by the quantum code,
then an uncorrectable error has occurred\footnote{Note, $H$ can
detect more errors than the quantum code can correct; the
quantum code stabilizer is formed from $H^{\perp}$ not $H$.}.

The success and crash frequencies provided by the program are
interpreted as follows. Let $n_f(Q)$ and $n_s(Q)$ be the
number of runs in which the quantum computer crashed
at step $Q$, and the number of runs in which the computer
remained successful at step $Q$,
respectively. The probability that the computer crashes during
step $Q$, given that it has not crashed in steps $1$ to $Q-1$, is
then
 \beq
 p(Q) = \frac{n_f(Q)}{n_s(Q) + n_f(Q)} \end{equation}
With stochastic noise, this probability is expected to be
independent of $Q$ once initial transient effects have died away,
and this was found to be the case. The transient behaviour was
found to last a few logical steps, with the general form $p(Q)
\simeq \bar{p} (1 - (5/4) \exp(-Q/2))$ for $Q \ge 1$
where $\bar{p}$ is the
average $p(Q)$ for large $Q$. Hence it was sufficient to
continue each run to 10 logical steps, and take $\bar{p} \simeq
(p(7)+p(8)+p(9)+p(10))/4$. For each case, the simulation was
repeated until $n_f(Q=10)$ reached 100, so the statistical
uncertainty in $\bar{p}$ is expected to be approximately 5\%. The
random part of the variation in $\bar{p}$ which is visible in
figures 1--3 is consistent with this expectation. For $Q>10$, the
value of $F_Q$
can be estimated as
  \beq F_Q \simeq (1-\bar{p})^Q.
\end{equation}

The estimation method to be presented in section \ref{s:est} was
used to predict the best choice of parameters $r,r',r''$ in the
last two steps of the protocol, and the choice was confirmed by
repeated runs of the Monte-Carlo calculations. One expects $r'>1$
to be necessary so that the probability of accepting a wrong
syndrome is not linear in the noise rates. If was found that for
the $[[7,1,3]]$ code, $r'=2$ was optimal, and very similar results
were found for $r=2$ or 3, $r''=1$ or 2. For the $[[23,1,7]]$
code, $r=4,\;r'=r''=3$ gave the best results for low noise rates,
and $r=3,\;r'=r''=2$ for high noise rates.

\begin{figure}[ht]
\centerline{\resizebox{!}{68 mm}{\includegraphics{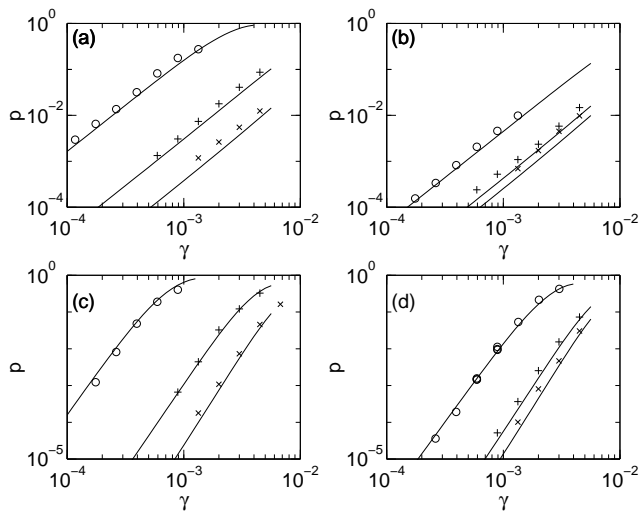}}}
\rule{0pt}{24 pt}
\caption{Results of numerical calculations of $\bar{p}$
(symbols) compared with the analytical
estimate (curves), at $\alpha n_{\rm rep} = \beta + r(1-\beta)$.
The symbols indicate $\epsilon = \gamma$ ($\circ$),
$\epsilon = \gamma/10$ ($+$), $\epsilon = \gamma/100$ ($\times$).
The calculation used $r=r'=r''=2$ for the $[[7,1,3]]$ code;
$r=4$, $r'=r''=3$ for the $[[23,1,7]]$ code.
(a): $[[7,1,3]]$, $t_m=25$;
(b): $[[7,1,3]]$, $t_m=1$;
(c): $[[23,1,7]]$, $t_m=25$;
(d): $[[23,1,7]]$, $t_m=1$.}
\end{figure}

\begin{figure}[ht]
\centerline{\resizebox{!}{68 mm}{\includegraphics{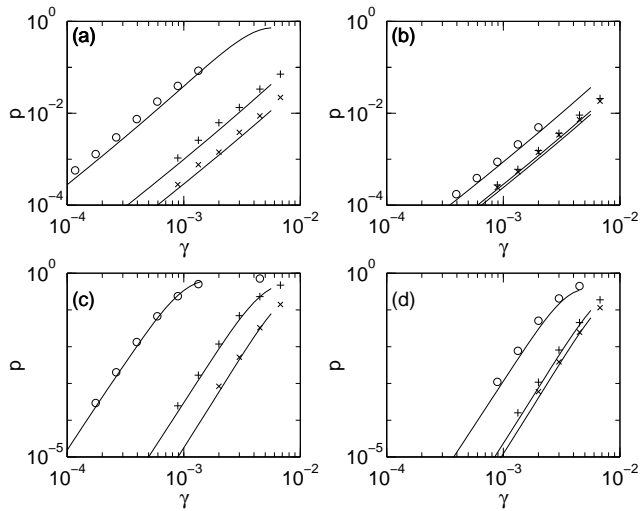}}}
\rule{0pt}{24 pt}
\caption{As figure 1, but for higher $n_{\rm rep}$, here
$\alpha n_{\rm rep} = 10(\beta + r(1-\beta))$.}
\end{figure}

\begin{figure}[ht]
\centerline{\resizebox{!}{35 mm}{\includegraphics{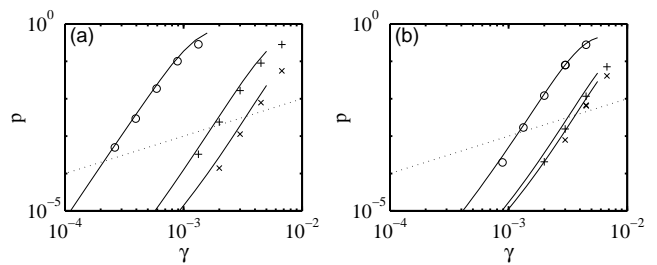}}}
\rule{0pt}{24 pt} \caption{As figure 2, but with reduced $r$
parameters, $r=3,\;r'=r''=2$ for the $[[23,1,7]]$ code. The dotted
lines show the `break-even' condition $\bar{p} = \gamma$ to facilitate a
rough estimate of the noise threshold from these results.}
\end{figure}

Figures 1--3 show example results of these calculations. In each
case the points indicate the results of the numerical
calculations and the lines show the prediction of the model to be
described in section \ref{s:est}.
The parameters
associated with the choice of code are
listed in table 1 of appendix A. The noise parameters were chosen to be
$\gamma_2 = \gamma_1 = \gamma_m = \gamma_p \equiv \gamma$, and results
for three values of $\epsilon/ \gamma$ are shown.
Changing
$\gamma_1$ and/or $\gamma_m$ by an order of magnitude while
leaving $\gamma_2$ fixed does not
have a large effect on the results, because the networks are
dominated by the 2-qubit gates. The value of
$n_{\rm rep}$ can be freely chosen, producing one route for the
trade-off between scale-up and noise-tolerance.
$n_{\rm rep}$ is accounted for in the
numerical calculations simply by adjusting the amount of memory
noise in the data bits occurring during each round of QEC. It was
convenient to treat the case that $n_{\rm rep}$ varies so that
the number of parallel corrections is the
same for all values of $\gamma$ and $\epsilon$ in a given set of
calculations.

The results in fig. 1
are for $\alpha n_{\rm rep}/( \beta + r (1-\beta)) = 1$ (i.e.
$n_{\rm rep} \simeq 1$) and those in fig. 2 are for
$\alpha n_{\rm rep}/( \beta + r (1-\beta)) = 10$
(i.e. $n_{\rm rep} \simeq 10$).
Fig. 3 shows the effect of reducing $r$: this is expected to
make matters worse
at very low $\gamma$, but better at higher $\gamma$. Comparison of figure
3 with figure 2 shows that a well-chosen reduction in $r$ makes possible
a useful increase in the noise threshold (see section
\ref{s:thresh}).

The comparison between the numerical results in figures 1--3 and the analytical
prediction will be discussed in section \ref{s:MCcomp}
after the analytical estimation method is described.

\section{Estimate of crash probability}  \label{s:est}

The numerical method permits the crash probability to be
calculated for small codes and high noise rate. A
quantum computer performing
a large computation will require lower noise rate and
larger codes which are able to correct more errors.
The Monte-Carlo simulation is too slow to be useful it that
regime.
In this section I present a general analysis of the QEC protocol
which permits an estimate of the crash probability to be made for
any code and noise rate. The analysis will also be useful
in order to understand
the best strategy for code concatenation, to be considered in
section \ref{s:thresh}.

The main route by which the quantum computer crashes is that
too many errors accumulate in the data block between one round of
correction and the next. These errors are either
caused directly there by noise in the data qubits and the gates which act
on them, or they are the result of error propagation from the
noisy ancillas. The fault-tolerant design of the QEC network
ensures that each failure can only cause one error
in any given data block, and more generally each set of $m$ failures can
only cause total error of weight $m$
in a data block. Let $g$ be the number of independent gate
failure locations which can result in 1 error in the data block,
and $s$ be the number of independent memory failure locations
which can result in 1 error in the data block, during a single
recovery. The probability that an unspecified error of
weight $m$ appears in the data is given to good approximation
by
\beq
B'(g,s,m,\gamma,\epsilon) \equiv \sum_{j=0}^m B(g,j,\gamma)
B(s,m-j,\epsilon)
\end{equation}
where $B$ is the binomial function defined in equation (\ref{B}).
The sum gives the probability of no gate failures and $m$ memory
failures, plus the probability of 1 gate failure and $m-1$ memory
failures, and so on up to $m$ gate failures and no memory
failures. It involves a slight miss-counting
since sometimes different failures have the same effect, so
some sets of $m$ failures produce an error of weight $< m$.
However, this miss-counting is not expected to give the main
limitation on the accuracy of the whole calculation, for the networks
under consideration.

An error is uncorrectable if it has a weight larger
than\footnote{There are correctable errors of higher weight, such
as members of the stabilizer, but these have negligible
probability compared to uncorrectable errors when the noise is
uncorrelated and good minimum distance codes are used.} $t$. In
the limit of small $\gamma$, $\epsilon$, the expression $\bar{p}
\simeq 2 B'(g,s,t+1,2 \gamma/3, 2 \epsilon/3)$ is a rough estimate
for the crash probability per block per recovery, and hence it is
only necessary to estimate $g$ and $s$ for the QEC network in
order to roughly estimate $\bar{p}$ for a given code. The factors
of $2/3$ account for the fact that of all the errors affecting any
given qubit, on average $2/3$ require $X$-correction and $2/3$
require $Z$-correction. This is true for errors of any weight
because they are caused by uncorrelated failures. For example, of
the 9 possible 2-qubit errors, 2 require $X$-correction of the 1st
qubit alone, 2 of the 2nd qubit alone, and 4 of both qubits: these
numbers are correctly given by the model as $9 \times (2/3) \times
(1-2/3)$ (twice) and $9 \times (2/3) \times 2/3$.
The overall factor 2 in $\bar{p}$ is because both the $X$ error and the $Z$
error must be correctable.

\begin{figure}[ht]
\centerline{\resizebox{!}{4 cm}{\includegraphics{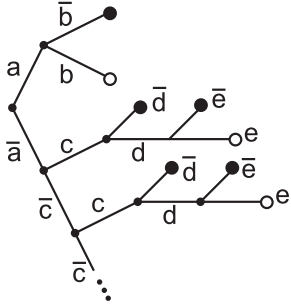}}}
\rule{0pt}{24 pt}
\caption{Probability tree to aid the calculation of $\bar{p}$.
The branches are labeled as follows.
a: First syndrome extracted is zero, b: the single syndrome
extraction left a correctable error, c: $r'$ of the most
recently extracted syndromes are found
to agree, d: the accepted syndrome is right, e: the multiple
syndrome extraction left a correctable error. The crash
probability is the sum of the probabilities of the branches
terminated by filled circles.}
\end{figure}

For a more precise estimate of $\bar{p}$, the protocol
must be analyzed more fully. A more complete analysis is indicated in
figure 4, which gives a probability tree for the full protocol.
I assume the quantum computer crashes not only when an
uncorrectable error occurs, but
also when a sufficiently bad syndrome is accepted (the latter
is discussed further in appendix B). I take into account
the fact that the values of $g$ and $s$ will depend on how many
syndromes have been extracted before an acceptable one is found
allowing a correction to take place. I will use the word {\em
recovery} to mean one attempt to get a consistent syndrome for each type of
error (which will involve either $1$ or $r$ or $r''$ syndrome
extractions for each type of error) followed by the corrections
which take place if sufficient agreement among syndromes is found
in step 8 of the protocol. The word {\em correction} will now refer to
the last stage of recovery only. Thus for any given data block
sometimes several recoveries have to take place before a correction
can be applied.

Consider the recovery of a single data block.
I will consider just the $X$-errors in the data block, and the
$X$-syndrome, bearing in mind that $X$ errors in the data are produced
partly by the network which extracts $Z$-syndromes.
The complete crash probability of the computer
per recovery per data block is assumed to be twice
the crash probability associated with the $X$-error recovery
of this single block.

Let $P_{Z\rm a}$ be the probability a verified ancilla, i.e. one
that was deemed `good' in steps 3 and 4 of the protocol, has one
or more $Z$-errors, so that it will produce an incorrect syndrome
for the data block.
\begin{widetext}
\begin{eqnarray}
P_{Z\rm a} &\simeq& \sum_{j=1}^n
B'\left( \frac{1}{2} N_{GV} + n ( 1 + \gamma_1/\gamma_2 +
\gamma_m/\gamma_2 ), \;
\frac{1}{2} N_{h} + t_m n,\; j,\; \frac{2}{3} \gamma_2,
\;  \frac{2}{3} \epsilon \right)                     \\
 & \simeq & \frac{1}{3} \gamma_2 N_{GV} + \frac{2}{3}\left( \gamma_2 + \gamma_1 + \gamma_m \right) n
 +  \frac{1}{3} \epsilon N_{h}  + \frac{2}{3} \epsilon t_m n
                         \label{PZa}
\end{eqnarray}
\end{widetext}
where $N_{GV} = 2 N_A + (n+k)/2$
is the total number of gates in the combined
$G$ and $V$ networks; it is dominated by the $2 N_A$ term, where
$N_A$ is the number of 1's in the $A$ part of the check matrix
$H=(I\, A)$ from which both the $G$ and the $V$ networks are
obtained; $N_{h}$ is the number of `holes' in the $G$ and $V$
networks, that is, the number of locations in space and time where
a qubit is resting and so experiences memory- but not
gate-failure:
\begin{eqnarray*}
N_{h} &=&
\left\{ w n - 2 N_A + 3(n-k)/2 \right\}  \\
&&  \mbox{} +
\left\{ w \left(n + (n+k)/2 \right) - 2 N_A + (n-k)/2 \right\}
\end{eqnarray*}
The parameter $w$ was discussed after equation (\ref{C}); the
first line is the number of holes in $G$, the second is the number
of holes in $V$.

The factors of $2/3$ and $1/2$ in (\ref{PZa}) account
for the fact that of all the failures occurring, some cause purely
$X$ error in the ancilla which does not cause a wrong syndrome,
and most of those that cause $Y$ error result in the
ancilla failing the verification, so they do not affect `good' ancillas.
The terms involving $\gamma_1$ and $\gamma_m$ are the
contributions from failure of the final Hadamard gates and measurements
of the ancilla bits. The further term involving $\gamma_2$ is from
the controlled gates connecting ancilla to data.
Failures of the preparation of ancilla bits in $\ket{0}$ do not
contribute to $P_{Z\rm a}$ because $\ket{0}$ is an eigenstate of
$Z$. The term involving $t_m$ is the contribution from memory failure in the
ancilla during the time taken for the verification bits to be
measured. (Equation (\ref{PZa}) is discussed further in appendix B).

The fraction $\alpha$ defined in step 4 of the protocol is
given approximately by
\beq
\alpha \simeq 1 - \frac{2}{3} \left( N_{GV}\gamma_2 + n \gamma_p + N_{h} \epsilon \right) .
\eeq
This is one minus the probability that a failure of type $X$ or $Y$
occurs in the $G$ and $V$ networks, since almost all such failures
are detected by the verification.

In what follows, I will be calculating probabilities for $X$
errors to be present in the data block when the $X$-syndrome
extraction is performed. These $X$ errors have three origins: the
gates associated with the logical operation which evolves the
logical quantum computation; the $^C\! Z$ gates which link an
ancilla or ancillas to the data block for $X$-syndrome extraction;
and the network for the preceding $Z$-syndrome extraction
(including error propagation from those ancillas to the data). In
a given recovery either 1 or $r$ or $r''$ extractions of each type
take place, I calculate the probability of each case and deduce
the average effect.

Let $g(r_X, r_Z)$ be the number of independent gate failures
resulting in an $X$ error in the data when a network accomplishing
$r_X$ $X$-syndrome extractions and $r_Z$ $Z$-syndrome extractions
is applied. For the same network, let $s(r_Z)$ be the number of
independent space-time locations of memory failures that result in
an $X$ error in the data. $s$ does not depend on $r_X$ because
propagation from the ancillas used for $X$-syndrome extraction
produces $Z$ not $X$ errors in the data.
\begin{eqnarray}
\!\!\!\!\!\!\!\! g(r_X, r_Z) &\simeq& n (1 + r_X + (1 + \mu t) r_Z),   \label{gr} \\
s(r_Z) &\simeq&  n \left(t_R + (\nu t  + t_m ) r_Z \right),
\label{sr}
\\ \mbox{where} \;\, t_R &=& (2 w + 1 + 2 t_m)
\frac{\beta + r(1-\beta)} {\alpha n_{\rm rep}}
\end{eqnarray}
and $\mu,\nu$ are constants of order 1 to be determined. The
fraction $\beta$ was defined in step 6 of the protocol; $t_R$ is
the time the data bits `rest' between successive recoveries. The
estimates of $g$ and $s$ are the most important to get right,
because they lead directly to the probability of uncorrectable
errors in the data.
In the expression (\ref{gr}) for $g$
the first term is caused by the $n$ elementary gates of a single
transversally-applied logic gate which may be present between recoveries in
the protocol adopted, the second term accounts for failure of the
transversal $\C\!Z$ gates connecting ancillas to the data block to
extract $X$-syndromes, and the third term the effect of the
$Z$-syndrome extractions. In the last case only, error propagation
from the ancilla causes $X$ errors in the data. These errors are
caused by failure of the last gates in the $V$ network; their
effect is estimated by the term $n \mu t r_Z$ in $g$ by the
following reasoning. The last gate of $V$ to act on each ancilla
bit can leave an $X$-error there which is not detected by the
verification bits; most pairs of gate failures from the last or
the penultimate set to act on each ancilla bit can leave
undetected single or double $X$-errors; triples of gate failures
from a still larger set can go undetected, and so on. This means
that the distribution of undetected ancilla errors caused by
failures in $V$ is not binomial: the number of failure locations
which can contribute to an order-$m$ failure is not independent of
$m$ but increases approximately linearly with $m$. I can
nevertheless use a binomial as an approximation to the true
distribution, as long as I make the approximation sufficiently
accurate for the most important probability I wish to calculate,
which is the probability of uncorrectable error in the data block.
For a $t$-error correcting code, this is the probability of
order-$(t+1)$ failures. The term $n \mu t r_Z$ in $g$, and a
similar term in $s$, approximately counts the relevant locations,
where the constants $\mu$ and $\nu$ were found by fitting the
theory to the numerical results, see figures 1--3 and section
\ref{s:MCcomp}. The values $\mu \simeq 0.35$, $\nu \simeq 1$ were
found to give the best fit.

Note that, as remarked in section \ref{s:protocol}, improving the
fidelity of the ancillas can only slightly reduce $g$ because it
can only reduce $\mu$ to a minimum of 0, and it can only allow a
slight reduction in the syndrome repetition parameters $r,r',r''$.

The first term in the expression for $s$ (equation (\ref{sr}))
accounts for the memory noise in the data block during the time
$t_R$ which has to pass between successive recoveries. $t_R$ can
be reduced by increasing $n_{\rm rep}$. If $t_R < (t_m +
r)$ then the syndromes for the next recovery are extracted before
the measurement of the current ones can be completed. However, as
long as the classical processing of the syndrome information takes
this into account, it need not be a problem. The rest of
(\ref{sr}) accounts for the memory noise in the ancillas which is
not detected by the verification and can propagate to the data.
The term $n \nu t r_Z$ follows from an argument similar to the one
just given for $g$, and the other term accounts for the period of
waiting while the verification bits are measured, which has to be
completed before the ancilla is coupled to the data (if it is
found to be good).

The fraction $\beta$ defined in step 6 of the protocol is equal to
the probability $P_0$ that the data block has no $X$ errors,
multiplied by the probability $(1-P_{Z\rm a})$ that this fact is
indicated correctly by the first syndrome extracted. I estimate
\begin{eqnarray}
P_0 &\simeq& \beta
B' \! \left(g(1,1),s(1), 0, \frac{2}{3} \gamma_2,\frac{2}{3} \epsilon
\right)  \nonumber \\
&+& \!\!\! (1 - \beta)
B' \! \left(g(1,r),s(r), 0, \frac{2}{3} \gamma_2,\frac{2}{3} \epsilon
\right) \! ,  \label{P0} \\
\beta &=& P_0 (1-P_{Z\rm a}).  \label{beta}
\end{eqnarray}
The reasoning is that since the last $X$-error correction, the
$Z$-syndrome extraction network required either 1 or $r$
syndromes, with probabilities $\beta$ and $(1-\beta)$
respectively, and only a single $X$-syndrome extraction has been
undertaken so far because we are at step 6 of the protocol. Note
that for equation (\ref{P0})
I have assumed that whenever the first syndrome is non-zero,
$r$ are extracted, which results in a slight underestimate of
$\beta$ since in fact sometimes $r'' < r$ are extracted. Also,
I ignore the variation of $\beta$ from one recovery to another. The
imprecision associated with these simplifications is small
compared to the imprecision of the whole calculation. Equations
(\ref{sr})--(\ref{beta}) are circular, but enable $\beta$ to be
found by iteration, starting from a value in the range $0 < \beta
< 1$.

Let ${\cal P}_1(r_X)$ be the probability that an uncorrectable error
accumulates in the data when $r_X$ $X$-syndromes, and either
1 or $r$ $Z$-syndromes, are extracted in
a single recovery attempt. I take an error of any weight above $t$
to be uncorrectable, so

{\samepage
\begin{eqnarray}
{\cal P}_1(r_X) &\simeq& \!\! \sum_{m=t+1}^{n}
\beta
B'\! \left(g(r_X,1),s(1), m, \frac{2}{3} \gamma_2,\frac{2}{3} \epsilon \right)
\nonumber \\
\lefteqn{\!\!\!\!\! +  (1 - \beta)
B'\! \left(g(r_X,r),s(r), m, \frac{2}{3} \gamma_2,\frac{2}{3} \epsilon
\right)\! .}     \label{calP1}
\end{eqnarray}
}
It is found that for a viable computer (i.e. $\bar{p} \ll 1$)
this is the largest contribution to the overall
crash probability $\bar{p}$.

Let $P_{\rm agree}(j)$ be the probability that in step 8 of the
protocol sufficient syndromes are found to agree for correction to
be completed, where $j$ is the number of
successive recoveries since the last time an $X$-syndrome was
accepted for the block in question, so that an $X$-error correction took
place.
I argue that agreement is found whenever $r'$ or more good syndromes
have been prepared without $Z$ error, hence
\beq
P_{\rm agree}(1) \simeq \sum_{m=r'}^r B\left(r,m,1-P_{Z\rm a}\right)
\eeq
and, using the protocol as in steps 7(b), 8(b),
\beq
P_{\rm agree}(j>1) \simeq \sum_{m=r'}^{r+r''}
B\left(r+r'',m, 1-P_{Z\rm a}\right).
\eeq
(This estimate breaks down at $r''=0$ but I always require $r'' \ge 1$.)

Let $P_{\rm ws}$ be the probability of a crash caused by
a group of $r'$ syndromes conspiring to agree on a syndrome, even
though they are all wrong, which would result in the wrong `correction'
being made to the data. I estimate
\beq
P_{\rm ws}  \simeq  N_{GV} (\gamma_2/3)^{r'}
                       + N_{h} (\epsilon/3)^{r'}. \label{Pwrong}
\eeq
$P_{\rm ws}$ is much smaller than $(P_{Z\rm a})^{r'}$
because to accept a wrong syndrome it is necessary that the {\em
same} error in the ancilla happens in $r'$ independent ancilla
preparations. Any single $Z$-failure will cause the ancilla to be
in a state of non-zero syndrome. Since there are many more
possible syndromes than individual failure locations in the
ancilla preparation network, it is rare that two different failure
locations give rise to the same final error in the ancilla. Therefore the
probability of obtaining an ancilla state of the same non-zero syndrome
in $r'$ independent preparations is, to lowest order
in $\gamma, \epsilon$, the probability that the same failure
happens in all the preparations.
This is approximately $(\gamma/3)^{r'}$
multiplied by the number of different gate failure locations,
plus a similar term accounting for memory failure.
The factors $1/3$ appear because almost all failures which produce
$Y$ errors are detected by the verification, so do not affect good ancillas,
and those which produce $X$ errors do not produce a wrong syndrome.
Note, (\ref{Pwrong}) does not include terms for the noise in the
gates connecting ancilla to data, nor the memory noise while the
verification bits are measured, nor noise in the ancilla
measurement. This is because noise at these locations causes
predominantly single-bit errors in the ancilla, and these are
almost harmless---see appendix B---the further contribution to
$P_{\rm ws}$ is negligible when $(t_m \epsilon)^2 \ll \gamma_2$
and $\gamma_m^2 \ll \gamma_2$.

It is found that for small codes and/or high noise rate, smaller values
of $r,r',r''$ are better, in order to reduce $\cal P$; for
large codes and/or low noise rate,
higher values of $r,r',r''$ are better, in order to reduce $P_{\rm ws}$ and
to keep $P_{\rm agree}$ sufficiently large.
Once $r'$ is large enough, the value of $\bar{p}$ is insensitive
to $P_{\rm ws}$ because it is dominated by the other
terms.

I can now calculate $\bar{p}$, using the probability tree shown
in figure 4 as a guide:
\begin{eqnarray}
\bar{p}(C,\{ \gamma_i \},\epsilon)
&\simeq& 2 \left\{ \beta {\cal P}_1(1) + (1 - \beta)
\left[ P_{\rm agree}(1)   \right. \right. \nonumber \\
\lefteqn{ \left. \left.
\left( P_{\rm ws} + (1-P_{\rm ws}) {\cal P}_1(r)
\right) + S \right] \right\} }
  \label{pbar}
\end{eqnarray}
where $C$ refers to the set of parameters $\{ n,k,t,w,N_A,r,r',r'',t_m,n_{\rm
rep} \}$ and $\{ \gamma_i \} = \{\gamma_1, \gamma_2, \gamma_p,
\gamma_m \}$.
$S$ is the sum of the probabilities associated with
the lower branches of the tree, when the first recovery attempt
gave no consistent syndrome. To calculate these,
rather than keeping account of all the possibilities, I use an
average for the number of $Z$-syndrome extractions which
take place in each recovery. This average is
\beq
\bar{r} \simeq \beta + (1-\beta) (P_{\rm agree}(1) r +
(1-P_{\rm agree}(1))r''). \eeq
I then have for the probability of uncorrectable $X$-error after a
total of $j > 1$ recovery attempts since the last correction:
\begin{widetext}
\beq
{\cal P}_j \simeq  \sum_{m=t+1}^{n}
B'\left(g(r + (j-1)r'',j \bar{r}),s(j \bar{r}), m, \frac{2}{3} \gamma_2,
\frac{2}{3} \epsilon \right).
\label{calPj}  \eeq
and
\beq
S \simeq \sum_{j=2}^{\infty}
\left\{ \prod_{i=1}^{j-1} \left( 1 - P_{\rm agree}(i) \right)
\right\} P_{\rm agree}(j)
\left( P_{\rm ws} + (1-P_{\rm ws}) {\cal P}_j \right)/j.
\label{S}
\end{equation}
\end{widetext}
The final division by $j$ accounts for the fact that I am
calculating an average crash probability per attempt at recovery.
The logical quantum computation continues whether or not any one
recovery attempt gave a consistent syndrome.

\subsection{Illustrative example}

To illustrate the main features of the calculation, consider for
example the $[[127,43,13]]$ BCH code, for parameter values
$\gamma=10^{-4}$, $\epsilon = 10^{-6}$, $n_{\rm rep}=2.5$, $t_m =
25$, and we choose $r=5,\; r'=4,\; r''=3$. The code has $w=47$,
$N_A = 1802$ (see table 1) giving $N_{GV} = 3689$, $N_h = 8893$.
In each time step approximately $N_A/w \simeq 38$ gates act in
parallel on each ancilla during preparation and verification; the
recovery time is $t_R = 143$ time steps.

These parameter values give $\alpha \simeq 0.74$, $\beta \simeq
0.8$. Suppose the computer consists of 10 blocks. Of the 25
ancillas prepared for $X$-recovery, on average $25 \alpha \simeq
18$ are found to pass the verification. When the first syndrome is
extracted for each block, $10 \beta \simeq 8$ are found to be
zero, 2 non-zero. For $r=5$, a further 4 syndromes are extracted
from each of the two blocks needing further attention, this uses
up the remaining 8 ancillas which passed verification. The ancilla
error probability is $P_{Z \rm a} \simeq 0.1$ and the probability
a data block has no errors is $P_0 \simeq 0.9$, therefore the
typical situation is that one of the two blocks being recovered is
found in fact to be free of errors (its first reported non-zero
syndrome was wrong, caused by a $Z$-error in the ancilla
preparation) while the other has a correctable error. The
probability that 4 of the 5 syndromes agree is $P_{\rm agree}
\simeq 0.8$ so the error is usually identified first time, but
occasionally this must await the next recovery. In the latter case
the subsequent recovery of the block in question has $r+r''=8$
syndromes available, the probability that 4 of them are mutually
consistent is approximately $0.998$.

The probabilities of uncorrectable error in the data (branches
$\bar{\rm b}$ and $\bar{\rm d}$ in fig. 4) are
 \begin{eqnarray*}
{\cal P}_1(1) &\simeq&  3 \times 10^{-13} \beta +  5 \times 10^{-10} (1-\beta)
\; \simeq \; 3^{-11} \\
{\cal P}_1(r) &\simeq&  8 \times 10^{-12} \beta +  2 \times 10^{-9} (1-\beta)
\; \simeq \; 4 \times 10^{-10}
 \end{eqnarray*}
while the probability of accepting a wrong syndrome is $P_{\rm ws} \simeq 5 \times 10^{-15}$. The overall result is $\bar{p}
\simeq 3 \times 10^{-10}$. It is seen that the main contribution
to the crash probability comes from the occasions where repeated
syndrome extractions are required for both the $X$ and $Z$
recoveries, leading to too many errors in the data. On these
occasions the number of gate and memory failure locations is
$g(r,r) = 2540$, $s(r) \simeq 39000$ respectively, therefore gate
failure dominates when $\epsilon < \gamma/15$.

\subsection{Comparison of analytical estimate and Monte-Carlo simulation}
\label{s:MCcomp}

 The prediction given by equations
(\ref{PZa}-\ref{S}) is shown by the curves in figures 1-3. The
main feature of both the numerical results and the analytical
estimate is that $\bar{p}$ varies as $(\gamma + \mbox{const.}
\epsilon)^{t+1}$ in the limit of small $\gamma$, $\epsilon$. This
is seen for example in equation (\ref{calP1}), where a useful pair
of approximations is
\begin{eqnarray}
B'(g,s,m,\gamma,\epsilon) &\simeq& B\left(g,m,\gamma + s \epsilon/g
\right) \\
&\simeq&
\left( \frac{g \gamma + s \epsilon}{m/e} \right)^m
\end{eqnarray}
The first approximation is quite accurate in the regime under
consideration (small $\gamma$, $\epsilon$), while the second gives
the right order of magnitude; $e$ is the base of natural
logarithms and I used Stirling's formula to simplify $m!$ (even though $m$
is not large).

The values of the fitted parameters $\mu$ and $\nu$ were adjusted
to get the best fit between the curves and the Monte-Carlo `data'.
This implies that the curves must match the data in at least two
places. The fact that the curves also correctly show all the major
trends as a function of $\epsilon$, $\gamma$, $n_{\rm rep}$, $t_m$
and the code parameters is evidence that the analysis is
sufficiently complete to be useful. The analytical estimate
predicts that $\bar{p}$ falls slightly faster than $\gamma^{t+1}$
in the region $10^{-4} < \bar{p} < 10^{-2}$ because the mean
number of syndromes extracted is falling as $\beta$ increases
towards 1. The simulations are consistent with this but in the
absence of simulated points at $\bar{p} < 10^{-4}$ it was not
possible to confirm it thoroughly. The Monte-Carlo simulation was
too slow to explore the latter region (each point at $\bar{p}
\simeq 10^{-4}$ took many days to compute on a modern
workstation).

The agreement overall is good. The main (but still modest) discrepancy is
that in figures 1 and 2 the analysis underestimates the numerical
results, while in figure 3 it overestimates. In other words, the
effect of changing the $r$ parameters is greater in the numerical
simulation than in the analysis, in the region of large $\gamma$ and
small $\epsilon/\gamma$. By adjusting $\mu$ and $\nu$ it was possible
to get a
better fit either to figures 1 and 2, or to figure 3; the choice
shown ($\mu = 0.35,\; \nu = 1$) represents the best compromise.

There is a small systematic disagreement in gradient
for several of the sets of results,
especially in figure 1. This would be enough to cause a
disagreement in $\bar{p}$ by an order of magnitude if it persisted
to lower $\gamma$ values of order $\gamma = 10^{-4}$ (where
a direct comparison between numerical and analytical results
is not available). However, the analytical model always
produces the power-law behaviour $\bar{p} \propto \gamma^{t+1}$
at low $\gamma$ (as long as $r' \ge t$) so the disagreement
in gradient will not persist to low $\gamma$, and in any case
when $\bar{p} < 10^{-8}$ even an order of magnitude error
in $\bar{p}$ will only be a
relatively small effect in the results to be obtained from the
model in the rest of the paper.

The part of the analysis which can only be confirmed
to a limited extent by the simulations is the linear scaling with
$t$ of the terms $\mu t$ and $\nu t$ in (\ref{gr}), (\ref{sr}).
Simulations of more codes, especially codes correcting more errors,
would be necessary to give further information.

\subsection{Performance of a selection of codes}

The estimated crash probability (\ref{pbar}) was calculated for a
variety of codes with scale-up $N/K$ in the range 7 to 400.
The gate and measurement noise parameters were set to
$\gamma_1 = \gamma_2 = \gamma_m = \gamma_p \equiv \gamma$, and
$\bar{p}$ was calculated for several values of $\gamma$ with
$\epsilon = \gamma/100$ and with $\epsilon = \gamma/10$, at $t_m=25$, $n_{\rm rep}=1$.
The values
of $r,r',r''$ were adjusted to minimize $\bar{p}$ for each case.
To make a useful comparison, I then consider not $\bar{p}$
directly, but rather the number of qubit-gates $KQ$ which the
stabilized computer can achieve, allowing for the fact that codes
with $k=1$ allow slightly more efficient fault-tolerant gates than
codes with $k > 1$, c.f. section \ref{s:number}.
Using the method of `propagating the
gate through a teleportation' \cite{99:GottesmanB} only
approximately twice as many
recoveries per gate are needed when $k>1$ than when $k = 1$
\cite{99:SteaneB}, so $KQ = 1/\bar{p}\; (0.5/\bar{p})$ when $k=1$
($k>1$) respectively. The resulting values of $KQ$ are plotted in
figure 5 as a function of the scale-up for the code employed. The
codes themselves are identified in table 1 of appendix A.

\begin{figure}[ht]
\centerline{\resizebox{!}{110 mm}{\includegraphics{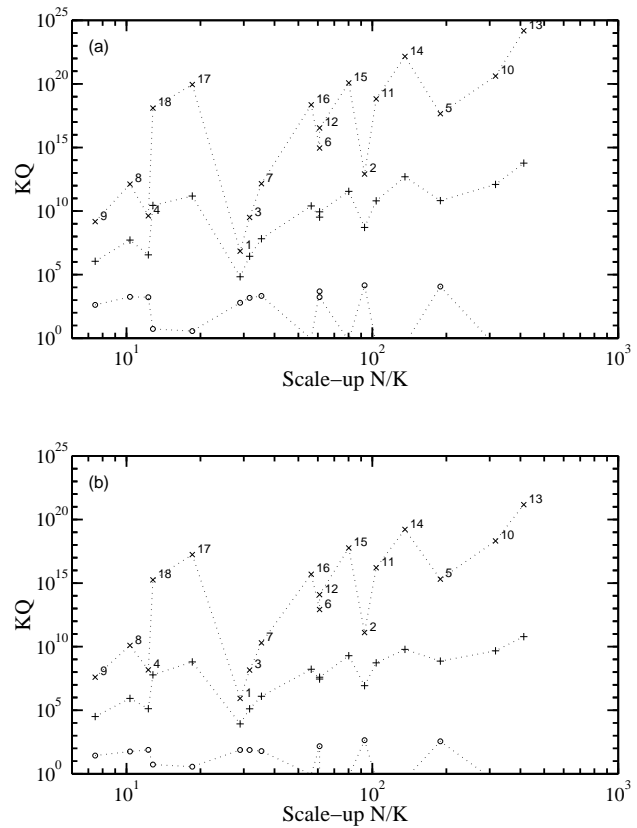}}}
\rule{0pt}{24 pt} \caption{Available $KQ$ for a variety of codes,
plotted against the scale-up $N/K$ of each code at $n_{\rm
rep}=1,\;t_m=25$. (a): $\epsilon=\gamma/100$, (b): $\epsilon =
\gamma/10$. The sets of results are at
$\gamma=10^{-5}$ ($\times$), $10^{-4}$ ($+$), $10^{-3}$ ($\circ$);
each code has a given
scale-up, the codes are indicated by an index number placed by
each point at $\gamma=10^{-5}$, which refers to the list in table
1. The dotted lines joining the points at given $\gamma$ are to
guide the eye.}
\end{figure}

Clearly at given values of $\gamma,\;\epsilon$ one would choose
the code of smallest scale-up which allows a desired $KQ$ to be
attained. The main conclusion to be drawn from figure 5 is that
for $\gamma \le 10^{-4}$, the efficient codes such as
$[[127,43,13]]$, $[[127,29,15]]$ perform well, but at higher noise
level other codes such as those numbered 1 to 9 in table 1 should
be adopted, at a cost in scale-up.

\section{Code concatenation}  \label{s:conc}

In order to get still smaller crash probabilities, and hence
larger algorithms, we need codes which can correct more errors.
Such codes exist, for example a $[[511, 241, 31]]$ BCH code, but
they of necessity involve more parity checks and therefore larger
networks to extract syndromes. The increase in $g$, $s$ and $P_{Z
\rm a}$ trades off against the increase in $t$, and as a result
these codes only become useful at low values of the failure rates,
$\gamma < 10^{-5}$.

Code concatenation enables this trade-off to be avoided, at the
cost of increased scale-up.

\subsection{Crash probability of concatenated code} \label{s:cpcc}

For stabilizing quantum memory, any code $C^i = [[n^i, k^i, d^i]]$ can
be concatenated with any other code $C^o = [[n^o, k^o, d^o]]$,
but for quantum algorithms the task of
constructing logical gates is rendered much more simple if $k^i =
1$ and both codes are CSS, therefore I restrict attention to this
case.
With $k^i=1$, $n^i n^o$ physical bits are used to
store $k^o$ logical bits.
$C^i$ is called the inner code, and $C^o$ the outer code, and their
combination is called the supercode.
The physical bits will be called `level-0' bits. Consider
$n^o$ groups containing $n^i$ physical qubits each.
To build the logical zero state of the concatenated code,
first prepare
each group of $n^i$ level-0 bits in the logical
zero of $C^i$ (e.g. by using a fault-tolerant measurement).
Each such block is then one `level-1' bit.
Next use transversal Hadamard and controlled-not
operations to evolve the $n^o$ level-1 bits into the logical zero
state of $C^o$. This network may or may not involve a level-1
recovery (i.e. recovery of the level-1 qubits in parallel, using
the inner code) before each transversal gate.

A concatenated code can be regarded in two ways. First, it can be
regarded as a single CSS code of parameters $[[n^i n^o, k^o,
d]]$ where $d = (d^i d^o + d^i + d^o -1)/2$. The methods described
in previous sections apply directly, the only change being in the
way the classical processor interprets the syndromes. Owing to the
code construction, to be uncorrectable an error must be composed of
more than $t^i = (d^i-1)/2$ single-bit errors in each of more
than $t^o = (d^o-1)/2$
sub-blocks. The probability for this is approximately
\beq
B\left(n^o,t^o+1, B(n^i,t^i+1,p)\right)
\eeq
where $p$ is the single-physical-bit error probability. The
equations (\ref{calP1}) and (\ref{calPj}) for $\cal P$ have to
be adjusted accordingly. This
first approach produces useful behaviour when the 7-bit and 23-bit
codes are combined once with themselves or each other, but for larger
codes the $G$ and $V$ networks become too large to allow recovery unless
the noise rates $\gamma$ and $\epsilon$ are very low.

The second way to use a concatenated code is to make more use of
its structure, by recovering the encoded level-1 qubits
inside the network which prepares level-2 ancillas. For example,
if after every gate in the level-2 network, a level-1 recovery is
applied to all level-1 qubits, then the overall behaviour is
described by the analysis given in section
\ref{s:est}, i.e. equations (\ref{PZa}) to (\ref{S}), applied to
the blocks of level-1 qubits. The gate and memory failure probabilities
$\gamma_1, \epsilon_1$ of
the level-1 qubits are estimated as the crash probability per
block per recovery associated with the inner code, i.e.
$\gamma_1 = \epsilon_1 = \bar{p}(C^i, \gamma, \epsilon)$, and
then the crash probability of the supercode is
$\bar{p} \simeq \bar{p}(C^o, \gamma_1, \epsilon_1)$.

We can improve matters further by distributing the inner
recoveries more intelligently. The main point is that it is better
not to recover resting qubits when this will make them more noisy.
To do
better, after the initial preparation of the level-1 qubits at the
beginning of the level-2 $G$ network, a level-1 recovery is
applied in parallel to all level-1 qubits, but thereafter
recover only non-resting level-1 qubits, just before a gate acts
on them, with one exception. The exception is the qubits in the
data block, which rest for a long time if $t^i_R \gg 1$, so these
qubits are given level-1 recoveries at the same rate as the qubits
in the ancilla. With this method, each level-1 qubit is recovered
on average once every $\eta \simeq 1 +
N^o_{h}/2 N^o_{GV} \simeq 2$ steps of the level-2
network. The effect can be estimated by
replacing the term $n^i t^i_R$ in equation (\ref{sr}) by $\eta n^i
t^i_R$ in the calculation of $\gamma_1 = \epsilon_1 =
\bar{p}(C^i,\gamma,\epsilon)$, and then for the calculation of
$\bar{p} = \bar{p}(C^o, \gamma_1, \epsilon_1)$ use the fact that
the memory noise in between gates of the outer $G$,$V$ networks has been
absorbed into $\gamma_1$, therefore set
\beq
N^o_{h} = 0, \;\;t^o_m = 1,   \label{No}
\end{equation}
and replace $n^o t^o_R$ by $n^o t^o_R/ \eta$. This more
intelligent placement of inner
recoveries was found to reduce $\bar{p}$ for all the cases plotted
in the figures.

A small saving on ancilla preparation can be obtained by
re-using the $n^o$ level-$1$ qubits of any ancilla which
failed its level-$2$ verification.

\begin{figure}[ht]
\centerline{\resizebox{!}{100 mm}{\includegraphics{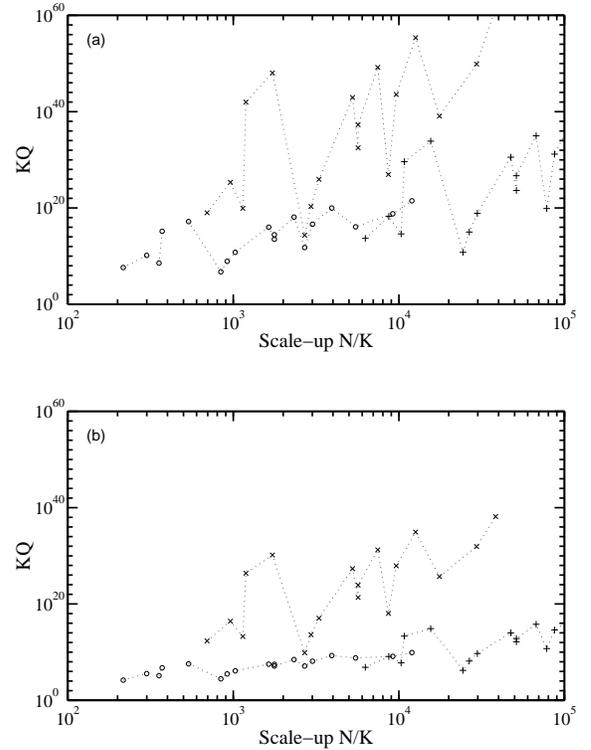}}}
\rule{0pt}{24 pt} \caption{Available $KQ$ verses overall scale-up
for codes concatenated with an inner $[[7,1,3]]$ Hamming code
($\circ$) or $[[23,1,7]]$ Golay code ($\times$) or concatenated
twice with the Hamming code ($+$). The results are given for the
same set of codes as in figure 5 and table 1. The case
$\gamma=10^{-4}$, $t_m=25$ is shown, for a computer with $n_{\rm
rep}=1$ for both inner and outer code. (a): $\epsilon=\gamma/100$,
(b): $\epsilon = \gamma/10$.}
\end{figure}

I show in figure 6 the effect of concatenating the codes of table
1 with the $7$-bit or $23$-bit code once, or with the $7$-bit code
twice, estimated
by the method just described, at $n_{\rm rep}=1$, $t_m = 25$.
The main conclusion is that the $23$-bit code is a better choice
than the $7$-bit code, since for given scale-up it permits the
higher $KQ$. It is clear that a great increase in $KQ$ is available
from the concatenation of the 23-bit code with another code.
In particular, the combination $C^i = [[23,1,7]]$, $C^o = [[127,29,15]]$
gives a very stable computer for scale-up around 1000.

\subsection{Threshold}  \label{s:thresh}

So far I have calculated the size of computation $KQ$ which can
be achieved for a given scale-up and given values of the failure
probabilities $\gamma, \epsilon$.
Further use of concatenation leads to the `threshold result',
which is the result that the situation $\bar{p} \rightarrow 0$ can
be obtained for values of $\gamma$, $\epsilon$ below a
threshold $\gamma_0$, $\epsilon_0$ which does not depend on $K$
and $Q$, assuming that the noise per qubit and per gate
is independent of the size of the computer and is stochastic
and uncorrelated, and sufficient parallel operation is available
in the computing device.
The threshold result may be proved by analyzing the case of a
particular code such as $[[7,1,3]]$, concatenated to arbitrarily
many levels. $\bar{p} \rightarrow 0$ is obtained when the crash
probability at each of the higher levels is less than that of the level below.

The protocol I have adopted is not guaranteed to be the absolute
optimal one, and in particular a protocol which had a higher scale-up
and allowed a slightly higher threshold may exist. However, the
protocol has been optimized in several ways, such as
minimizing the number of gates which connect ancillas to data, and
minimizing the time to verify ancillas, therefore it is unlikely
that another protocol will offer significant increases in the
threshold, under the assumptions which have been made about
the noise and the timing.
After the first two levels,
$\bar{p}$ is $O(\gamma^{(t+1)^2})$, i.e. varying quickly with
$\gamma$, and therefore the threshold is insensitive to details of
the protocol at higher levels.

An estimate of the threshold is immediately available by using the
analysis described in section \ref{s:cpcc}, extended to many
levels. I use the analytical estimate
$\bar{p}_L = \bar{p}(C,\bar{p}_{L-1},\bar{p}_{L-1})$, employing the
adjustment given by (\ref{No}), for the second level,
and then higher levels are modeled by
taking $t_m=1$ without any adjustment to $N_{h}$.
The adjustment of $t_R^i$ by $\eta$ described just before
(\ref{No}) has negligible effect when $n_{\rm rep}$ is large so
does not affect the maximum possible threshold.

\begin{figure}[ht]
\centerline{\resizebox{!}{65 mm}{\includegraphics{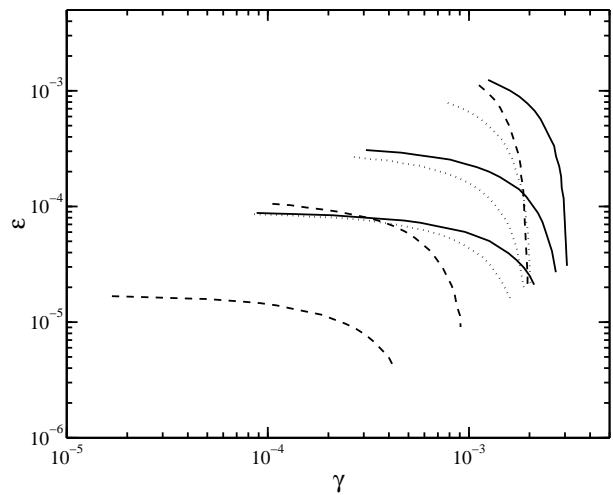}}}
\rule{0pt}{24 pt}
\caption{Threshold values $\gamma_0$ and $\epsilon_0$, for
$t_m=1$, 25 and 100. Values in the range
$0.01 \le \epsilon/\gamma \le 1$ are plotted.
Full curves: concatenated $[[23,1,7]]$, dashed curves:
concatenated $[[7,1,3]]$, dotted curves: concatenated $[[47,1,11]]$.
The highest value of $t_m$ produces the
lowest curve of each triplet.}
\end{figure}

The calculated threshold is shown in figure 7 for the codes
$[[7,1,3]]$ Hamming, $[[23,1,7]]$ Golay, and $[[47,1,11]$
quadratic residue, for three values of $t_m$ at the innermost
level, and for $0.01 \le \epsilon/\gamma \le 1$. It is seen that
if the measurements are fast ($t_m=1$), the two smaller codes give
a similar threshold, that of the Golay code being somewhat higher.
For the more physically realistic case of slow measurements ($t_m
\gg 1$), the Golay code offers a threshold higher than that of
the Hamming code by a factor
2 to 5. The Golay threshold values are in the region of $10^{-3}$
when $t_m = 1$ or $\epsilon \ll \gamma$, falling to $\sim 10^{-4}$
when $t_m \gg 1$ and $\epsilon = \gamma$.

It is instructive to compare this threshold calculation with previous
estimates. Previous calculations have all adopted the concatenated $[[7,1,3]]$
code rather than the Golay code, and typically no statement is
made about the measurement time $t_m$, but a value $t_m = 1$ is
implied. Gottesman and Preskill \cite{Th:Gottesman,98:Preskill}
quoted as a `conservative estimate' $\epsilon_0 = 10^{-5}$ when
memory noise dominates, and $\gamma_0 = 10^{-4}$ when memory noise
is negligible; in subsequent work the same authors derived
approximate values $6 \times 10^{-4}$ for both parameters,
with the caveat that these were overestimates, but that the true value
would exceed $10^{-4}$ \cite{97:Preskill}.
Aharonov and Ben-Or \cite{98:Aharonov} found $10^{-6}$ in a model
where measurement and classical computing is avoided, where one
expects a lower threshold. Zalka \cite{99:Zalka} found $\epsilon_0
= 10^{-4}$ when memory noise dominates, and $\gamma_0 = 10^{-3}$
when memory noise is negligible. In his calculation Zalka assumed
many logical gates can take
place between recoveries without causing an avalanche of errors.
My values for the case of $t_m=1$
and the $[[7,1,3]]$ code are $\epsilon_0 = 1.3 \times 10^{-3}$ and $\gamma_0
= 3 \times 10^{-3}$, where recovery takes place after every logical
gate so that the avalanche is avoided. My values are significantly
higher than previously reported ones, especially $\epsilon_0$
which is two orders of magnitude larger than the early ``conservative
estimates'', and one order of magnitude larger than the estimate by Zalka,
despite the fact that I uphold a further constraint in
the requirement to recover every block after every logical gate.
This is a real improvement, not simply
a lack of precision in the estimates, because I have taken
advantage of the insights presented in \cite{02:SteaneA} which
speed the verification of ancillas and hence increase the
tolerance of memory noise.
Furthermore, by recognizing the
advantage of the Golay code, which is more important when
measurements are slow ($t_m \ll 1$), the present study
reveals an increase in the gate noise threshold $\gamma_0$
by an order of magnitude at $t_m = 100$, compared to what would be
the case for methods previously studied, and an increase in the
memory noise threshold $\epsilon_0$ by between one and two orders
of magnitude, representing the improvement offered by the combination
of faster verification combined with better coding.

I estimate the uncertainty of my threshold estimates
to be approximately a factor 1.5; this
is simply a judgement based on the degree
of change in the results which was observed as refinements were added
to the calculation.

\section{$KQ$ surface and discussion}

I now bring together all the methods discussed above in order to
find the largest algorithm-size
$KQ$ which can succeed, as a function of the noise rate and the
scale-up, maximized over all the codes and parameter choices. This
is done by allowing
$n_{\rm rep}$ to take on a range of values, and calculating
$KQ$ and the scale-up for each code (including concatenated ones),
using whichever values of $r,r',r''$ give the highest $KQ$.
The values of $\log_{10} (N/K)$ are
then binned at 5 bins per decade, and the maximum value of $KQ$
in each bin is noted. This leads to a surface of $KQ$ as a function
of scale-up and noise rate. The surface is plotted (on a logarithmic
scale) in figure 8 for $\epsilon = \gamma / 100$, $t_m = 25$. The
optimal values of the $r$ parameters are listed in table 1 for the case
$\gamma = 100\, \epsilon = 10^{-4}$, $t_m = 25$.
Figure 9 shows lines of constant $\gamma$ and contours of constant $KQ$,
for the cases $\epsilon/\gamma =0.01$ and 1, at $t_m = 25$.
Figure 10 shows lines
of constant $\gamma$ and contours of constant $KQ$,
for the cases $t_m = 1$ and $100$, at $\epsilon / \gamma = 1$.

\begin{figure}[ht]
\centerline{\resizebox{!}{6 cm}{\includegraphics{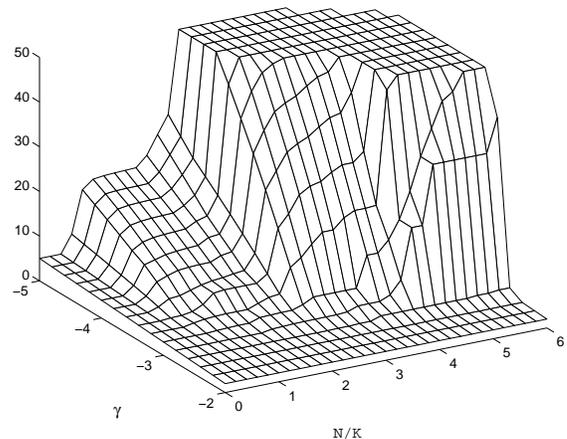}}}
\rule{0pt}{24 pt} \caption{Surface of maximum algorithm size $KQ$
as a function of $\gamma$ and scale-up $N/K$, at $\epsilon =
\gamma/100$ and $t_m = 25$. All the axes have logarithmic scales,
labeled in powers of 10. The surface has been truncated
at $KQ = 10^{50}$ to bring out the lower portions.}
\end{figure}

\begin{figure}[ht]
\centerline{\resizebox{!}{6.7 cm}{\includegraphics{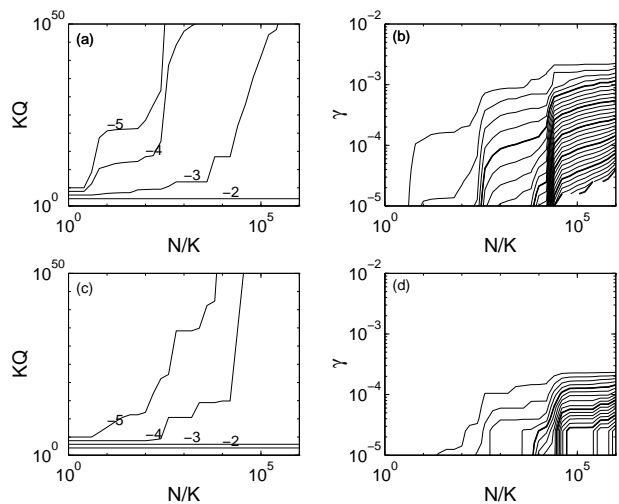}}}
\rule{0pt}{24 pt} \caption{(a),(c): Lines of constant $\gamma$,
for $\log_{10} \gamma = -2,-3,-4,-5$; (b),(d):
contours of constant $KQ$. (a), (b) for the case plotted in figure 8,
which has $\epsilon = \gamma/100,\;t_m = 25$. (c), (d) for the
case $\epsilon = \gamma,\; t_m = 25$.
The contours are
at $10^{10}$, $10^{20}$, $10^{30}$, etc.; every 5th contour is
shown with a thicker line.
}
\end{figure}

\begin{figure}[ht]
\centerline{\resizebox{!}{6.7 cm}{\includegraphics{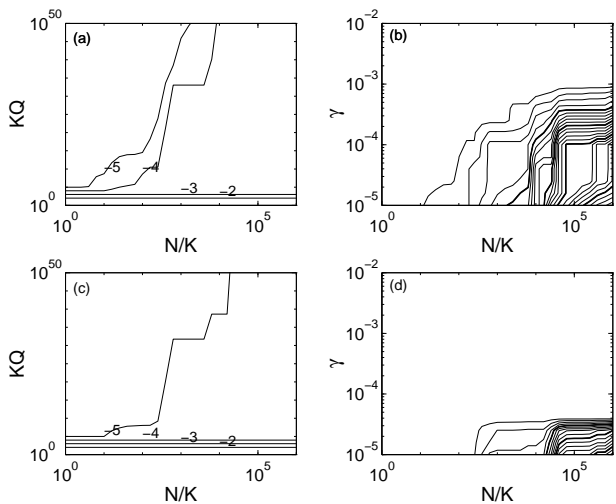}}}
\rule{0pt}{24 pt} \caption{As figure 9, but for the case
$\epsilon = \gamma$ and (a), (b): $t_m = 1$; (c), (d): $t_m = 100$.}
\end{figure}

The threshold result is indicated by the cliff
at $\gamma \simeq 2 \times 10^{-3}$
on the surface shown in figure 8, but this
cliff is not the only important feature of the surface.
Equally significant are the cliffs at $N/K \simeq 10$ and $N/K \simeq
1000$. The first indicates that large algorithms
($KQ \sim 10^{10}$) are possible for a modest scale-up once the
gate noise rate is $\le 10^{-4}$ (at $\epsilon = \gamma/100$), by using a
BCH code, and the
second cliff indicates that at the same noise rate
a scale-up of a few thousand allows
very large algorithms ($KQ \sim 10^{40}$), using the Golay code
concatenated once with a BCH code.

The really huge values of $KQ > 10^{20}$ should be interpreted
as an indication not that such large algorithms can necessarily
succeed, but rather that their failure will be for some other
reason not considered here, such as technical or environmental
problems causing correlated failure over many (e.g. hundreds of)
qubits.

Comparing figure 9b with figure 9d, it is seen that increasing
$\epsilon/\gamma$ from 0.01 to 1 at fixed $t_m$ has the effect
approximately of shifting the surface in the direction of smaller
$\gamma$ by an order of magnitude. Comparing figure 9d with
figure 10d, it is seen that an increase in $t_m$ from 25 to 100
at $\epsilon=\gamma$ has the effect
approximately of shifting the surface in the direction of smaller
$\gamma$ by almost another order of magnitude.

It should be re-emphasized that all the noise rates quoted apply
to non-local gates. They represent not the noise of a gate between
neighbouring qubits, or during the time for such a gate, but the
noise associated with a gate on qubits separated by some distance
which depends on the code and on the structure of the computer.
This distance has been discussed in \cite{02:SteaneB}, where it is
found that for certain reasonable choices of the layout of the computer,
the tolerated noise for a computer allowing only
nearest-neighbour coupling is expected to be about an order of
magnitude smaller than the tolerated noise for non-local gates
which has been given here. Also, in the case of only nearest-neighbour
couplings, at any given time-step in the $G,V$ networks, most
ancilla qubits which are said to be `resting' in the present
discussion will be involved in one or more swap operations. A
rough indication of the impact of this is obtained by letting the
memory noise parameter $\epsilon$ in the present discussion
include a contribution from the imprecision of swap gates.

A further assumption underlying all the results quoted is that
the computing device allows as much parallelism in its operation
as is logically possible for the networks considered. A reduction
in the available parallelism can be compensated to some extent by
a reduction in the memory noise.

\section{Conclusion}

The main results of this paper are figures 7 to 10, the set of
equations (\ref{PZa})--(\ref{No}), the network details set out in
table 1, and related insights such as the good performance of the
23-bit Golay code, and the role of the measurement time $t_m$.

The fundamental reason why the crash probabilities fall to such
low values is that uncorrelated and stochastic noise has the
special property that the likelihood of a cluster of failures
falls exponentially with the size of the cluster. There are two
main reasons why in practice the noise will not be like this:
qubit-qubit interactions and the technical details of the
machinery used to supply the gate operations. The former are
probably not too big a problem, since the strength of many-body
Hamiltonians typically falls very rapidly with the number of
bodies (see comments in section \ref{s:noise}). The latter can be
tackled by standard coding techniques such as random coding,
interleaving and concatenation. This implies that a study such as
the present one should be regarded merely as a starting point for the complete
structure of the computer. One
method to suppress correlations is to use a low-level encoding
such as $\ket{0}_L = \ket{0101} - \ket{0110} + \ket{1001} -
\ket{1010}$, $\ket{1}_L = \ket{0101} + \ket{0110} - \ket{1001} -
\ket{1010}$; this is a decoherence-free subspace for $ZZII$,
$IIZZ$ and $XXXX$ operators and so is unaffected by joint
$Z$-rotation of adjacent pairs of bits and joint $X$-rotation of
quadruplets of bits.

Further work in this area could address the following issues. Does
the error propagation directly between data blocks contribute
significantly to the crash probability, when recoveries are placed
in an optimal way as described in section 3? How well does the
simple noise model capture the main features of noise and
imprecision in particular physical examples? To what degree are
error processes present whose effects add coherently rather than
incoherently as assumed here? How much correlation and
non-stochastic behaviour is found in practice? Further numerical
simulations could be carried out on larger codes, mainly to test
equations (\ref{gr}), (\ref{sr}). The cost of moving information
around, and the transport distances required in QEC networks,
could be further analyzed so that noise the tolerance for
nearest-neighbour interactions can be calculated. The performance
of further encoding to suppress correlated noise could be studied.

\appendix
\section*{Appendix A: code construction}

\begin{table}
\begin{tabular}{rlcccccc}
code  & type & $n$ & $k$ & $d$ & $w$ & $N_A$ & r \\
number \\
0 & None     & 1 &  1 &  1 & -- & -- & -- \\
1 & Hamming  & 7 &  1 &  3 &  3 & 12 & 3 \\
2 & Golay    &23 &  1 &  7 & 11 & 77 & 4 \\
3 &  ,,      &21 &  3 &  5 &  7 & 63 & 4\\
4 & BCH      &31 & 11 &  5 & 15 & 122 & 4 \\
5 & QR       &47 &  1 & 11 & 15 & 281 & 5\\
6 & ,,       &45 &  3 &  9 & 15 & 255 & 4\\
7 & ,,       &43 &  5 &  7 & 15 & 229 & 4\\
8 & BCH      &63 & 27 &  7 & 27 & 350 & 4\\
9 & ,,       &63 & 39 &  5 & 27 & 328 & 4\\
10 & QR       &79 &  1 & 15 & 27 & 801 & 5\\
11 & ,,      & 77 &  3 & 13 & 27 & 759 & 5\\
12 & ,,      & 75 &  5 & 11 & 27 & 713 & 5\\
13 & QR      &103 &  1 & 19 & 31 & 1265 & 6\\
14 & ,,      &101 &  3 & 17 & 31 & 1215 & 5\\
15 & ,,      & 99 &  5 & 15 & 31 & 1165 & 5\\
16 & ,,      & 97 &  7 & 13 & 31 & 1119 & 5\\
17 & BCH     &127 & 29 & 15 & 47 & 1939 & 5\\
18 & ,,      &127 & 43 & 13 & 47 & 1802 & 5
\end{tabular}
\caption{Parameters of codes considered in the text. The code
constructions are outlines in appendix A. The parameters $w$ and
$N_A$ are the maximum weight of a row or column of the latin
rectangle for $A$, and the number of $1$'s in $A$, respectively.
The number of gates which act in parallel in most time-steps of
the generation or verification network of a given ancilla is
$N_A/w$. The final column gives the value of $r$ which was found
to be optimal when $\gamma = 100 \epsilon = 10^{-4}$, $t_m=25$,
with $r-1=r'=r''+1$.}
\end{table}

Table 1 lists the parameters of the codes considered in the text.
The values of $[[n,k,d]]$ are readily available from standard
texts such as \cite{Bk:Macwilliams}, but the values of $w$ and $N_A$
have to be obtained by examining the check matrices of the
classical codes. These were created using standard methods, see
\cite{Bk:Macwilliams} chapters 7, 9, 16.
The parity check matrix of a $[n=2^m - 1,k_c,d]$ classical BCH code
was created by letting $f$ be equal to an $m$'th-order polynomial which
is a factor of $1 + x^n$ over GF$(2)$. The check matrix consists of a
matrix of powers of $f$, where each entry is replaced by a column
of $m$ bits giving the coefficients in the polynomial $f^j$.

For a quadratic residue code having $n$ a prime which is one less
than a multiple of 4, the coefficients $f_i$, $0 \le i \le n-1$,
are defined to be
0(1) if $i$ is a quadratic residue (non-residue) respectively,
modulo $n$, and $f_0 = 1$.
The generator matrix is equal to the $n \times n$ circulant
matrix $G_{ij} = f_{j-i}$.

Further codes listed were obtained by deleting two columns from the
generator matrix of the classical code, to go from $[[n,k,d]]$ to
$[[n-2,k+2,d-2]]$, see \cite{96:SteaneC,98:Calderbank}.

Once the check matrix or generator matrix was obtained, it was
converted into the $(I A)$ form, and then $w$ and $N_A$ could be
obtained.

\section*{Appendix B: ancilla preparation statistics}

The ancilla preparation and verification was studied using the
Monte-Carlo method described in section \ref{s:num}. The method was to use
the noisy $G$, $V$ and ancilla--data coupling
networks to
extract a single syndrome from a data block which
was prepared with no errors. This was repeated many times
(starting from an error-free data block each time) and
the various syndromes obtained were counted, for values of
$\gamma$ in the range $10^{-5} < \gamma < 10^{-2}$.

Under the conditions of the numerical experiment, the obtained syndrome
should be zero. The non-zero syndromes indicate the errors
produced by the preparation/verification network which were not
detected during verification, and the further errors produced by
the coupling of ancilla to data, the final Hadamard gates on the
ancilla, and the measurements.

Let the syndromes be $\{ s \}$. The program gives the
probabilities $P_s(\gamma,\epsilon,t_m)$ of obtaining each $s$,
for a range of values of $\gamma$ at given $\epsilon/\gamma$ and
$t_m$. I fit each set of results to a power-law \beq P_s(\gamma) =
a_s \gamma^{c_s}  \label{Ps} \eeq where the fitted parameters
$a_s$ and $c_s$ depend on $\epsilon/\gamma$, $t_m$ and the code
under consideration. Only values of $P_s$ less than $0.01$ were
included in the fit, in order to avoid the non-power-law
dependence when $P_s$ approaches 1. I can thus examine the
statistics of the ancilla preparation in some detail by examining
the set of coefficients $a_s$, $c_s$.

\begin{figure}[ht]
\centerline{\resizebox{!}{5 cm}{\includegraphics{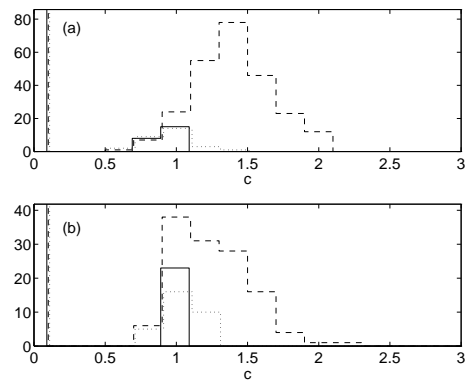}}}
\rule{0pt}{24 pt}
\caption{Distribution of $c_s$ values giving the power law
dependence of the probability of obtaining errors in the
prepared ancilla, as a function of
gate noise rate. The example given is for the Golay code at
$t_m=1$ and (a) $\epsilon = \gamma$, (b) $\epsilon = \gamma/100$.
Three histograms are plotted, showing the distribution for
errors of weight 1 (full line), 2 (dashed line) and 3 (dotted
line).}
\end{figure}

\begin{figure}[ht]
\centerline{\resizebox{!}{6 cm}{\includegraphics{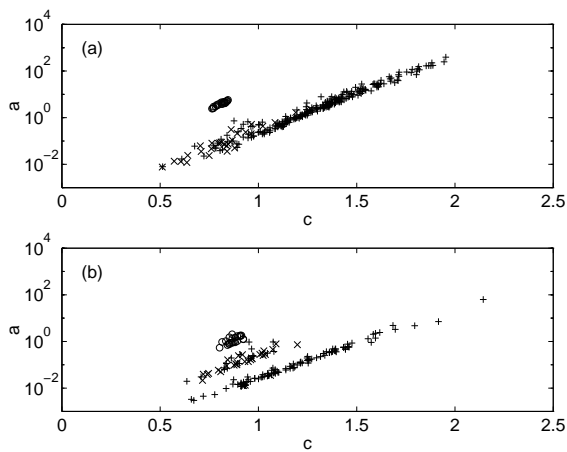}}}
\rule{0pt}{24 pt}
\caption{The fitted coefficients of equation (\protect\ref{Ps})
for all the ancilla errors observed in a large number of runs of the
Monte-Carlo simulation for the Golay code, at $t_m=1$.
(a) $\epsilon = \gamma$, (b) $\epsilon = \gamma/100$.
The symbols indicate the error weight
1 (o), 2 ($+$) or 3 ($\times$).}
\end{figure}

Figure 11 shows histograms giving the distribution of $c_s$ values
in the case of the Golay code, for $t_m=1$. The obtained syndromes
are first divided into sets, defined by the weight of the coset
leaders. I give the histogram for each set. This shows the
power-law dependence for preparing an ancilla with an error of
each weight. For
weight 1, I obtain $c_s = 1$, as expected: the
most likely cause of a single error is a single failure. For
weight 2, $c_s$ falls between 1 and 2, indicating that the most
likely weight-2 errors are caused by single failures or double
failures in roughly equal proportions. For weight 3, most $c_s$
values are close to 1, indicating that those weight-3 errors which
can be produced by a single failure are the most likely ones to
occur.

Figure 12 shows plots of $a_s$ verses $c_s$. There are two features
which stand out. First, there is a
correlation between $a_s$ and $c_s$, for the syndromes of each
error weight, such that $a_s$
increases by a factor of order $10^3$ when $c_s$ increases by
1. This means that the $P_s$ having small $c_s$ will
be more likely than those having large $c_s$ when
$\gamma < 10^{-3}$, which is the regime of interest. Therefore errors
(of whatever weight)
caused by single failures are the main ones I need to account for
in attempting to model the behaviour.

The second feature is that each weight-1 error is much
more likely to be produced than any individual weight-2 error.
This is to be expected: there are several single failure locations
which can produce a given weight-1 error in the ancilla, but only a
smaller number which can produce any given error of weight $>1$.

Note that at $\epsilon = \gamma/100$, fig. 12 shows that
most individual errors of weight 3 are
more likely than individual errors of weight 2, but fig. 11 shows
that there are fewer of them. This suggests the weight-3 errors
here are caused by gate failure in the preparation and
verification networks, while most of the weight-2 errors are
caused by memory failure.

\begin{table}
\begin{tabular}{ccccc|c}
$\epsilon/\gamma$  & $t_m$ & $a$ & $a'$ & $a+a'$ & $P_{Z \rm a}$ \\
1     & 1     & 196 & 36 & 214 & $243 \gamma$ \\
0.01  & 1     & 65  & 21 &  86 & $103 \gamma$ \\
1     & 25    & 509 & 54 & 563 & $609 \gamma$ \\
0.01  & 25    & 65  & 16 &  81 & $106 \gamma$
\end{tabular}
\caption{Linear fit coefficients $a$, $a'$ for the probabilities
of ancilla error of weight 1 ($a$, column 3) and weight $>1$
($a'$, column 4), obtained from the numerical calculations at
two values of $\epsilon/\gamma$ and of $t_m$. The final column
gives the probability of any ancilla error as estimated by
equation (\protect\ref{PZa}).}
\end{table}

A further statistic extracted from the calculations was the total
probability $P_w$ of obtaining any ancilla error of given weight
$w$, for weights between 1 and 4. This probability was fitted to a
power law (as a function of $\gamma$ at fixed $\epsilon/\gamma$).
The powers obtained were close to 1 (e.g. at $t_m=1$ the powers
were $0.97, 1.6, 1.4, 1.1$ for weights $1,2,3,4$ respectively,
when $\epsilon=\gamma$, and $0.93, 1.2, 1.1, 1.1$ when $\epsilon =
\gamma/100$.) In view of this and of the fact noted above, that
errors caused by single failures dominate the statistics when
$\gamma < 10^{-3}$, modeling these probabilities by a linear
dependence on $\gamma$ will capture the main features. Table 2
gives the fitted coefficient $a$ in the single-parameter linear
fit $P_{w=1} = a \gamma$, and $a'$ in the fit $P_{w=2} + P_{w=3} +
P_{w=4} = a' \gamma$, for two different values of
$\epsilon/\gamma$ and two different values of $t_m$. I expect the
net probability $P_{Z{\rm a}} = \sum_w P_w$ for
the ancilla to have some error
to be as given by equation (\ref{PZa}). The table shows that the
numerical results are fitted reasonably well by this model.

The other feature of the analysis presented in section \ref{s:est}
which I need to confirm is the value of $P_{\rm ws}$. This is
the probability of a crash caused by several successive ancillas
conspiring to agree on a wrong syndrome. Suppose the ancillas all
suffer from the same error vector $e$. When they couple to the
data, they pick up the error vector $d$ of the data bits, to give
a net error vector $e+d$. Assuming $e+d$ is correctable, the
correction applied to the data will be $e+d$, which leaves the
error $e$ in the data. If $e$ has small weight, this will not
cause a crash, and furthermore if $e$ has weight 1, it will only
add a further small contribution (scaling as the failure
rates raised to the power $r'$) to the
coefficient for single data errors, which is essentially harmless
\cite{02:SteaneA}. It follows that $P_{\rm ws}$ can be
estimated as the probability that $r'$ ancillas all have the same
error $e$ whose weight is greater than 1. Such errors are caused
mostly by that part of the $G$ and $V$ networks which takes place before
$V$ is completed, in which $X$ and $Y$ errors are mostly detected.
The probability for any
given $e$ is therefore either $\gamma/3$ or $\epsilon/3$, depending on
whether it was caused by a gate failure or a memory failure. The
number of different $e$ of weight $> 1$ that can be caused by a
single failure is overestimated by $N_{GV}$, ($N_{h}$), for those
$e$ caused by gate failure (memory failure) respectively, hence I
obtain the approximate value for $P_{\rm ws}$ given in equation
(\ref{Pwrong}). (For the Golay code the numerical study indicated
for that case the true numbers were $\simeq N_{GV}/4$ ($\simeq
N_h/4$).) If $t_m$ is sufficiently large, then the memory noise
while verification bits are measured will be such that a double
failure in this part of the network is as likely as a single
failure elsewhere; such a contribution can be neglected as long as
$(t_m \epsilon)^2 \ll \gamma_2$. If the measurement failure
probability $\gamma_m$ is sufficiently large then double
measurement failures will be significant; their contribution is
small as long as $\gamma_m^2 \ll \gamma_2$.

\acknowledgements
I would like to acknowledge helpful conversations with D. Gottesman.
This work was supported by the EPSRC and the Research Training and
Development and Human Potential Programs of the European Union.

\bibliographystyle{prsty}
\bibliography{quinforefs}

\begin{thebibliography}{10}

\bibitem{95:Shor}
P.~W. Shor, Phys. Rev. A {\bf 52},  R2493  (1995).

\bibitem{96:SteaneA}
A.~M. Steane, Phys. Rev. Lett. {\bf 77},  793  (1996).

\bibitem{96:Calderbank}
A.~R. Calderbank and P.~W. Shor, Phys. Rev. A {\bf 54},  1098  (1996).

\bibitem{96:SteaneB}
A.~M. Steane, Proc. Roy. Soc. Lond. A {\bf 452},  2551  (1996).

\bibitem{96:Shor}
P.~W. Shor,  in {\em Proc. 35th Annual Symposium on Fundamentals of Computer
  Science} (IEEE Press, Los Alamitos, 1996), pp.\ 56--65, quant-ph/9605011.

\bibitem{96:DiVincenzo}
D.~P. DiVincenzo and P.~W. Shor, Phys. Rev. Lett. {\bf 77},  3260  (1996).

\bibitem{98:Preskill}
J. Preskill, Proc. R. Soc. Lond. A {\bf 454},  385  (1998).

\bibitem{98:GottesmanA}
D. Gottesman, Physical Review A {\bf 57},  127  (1998), quant-ph/9702029.

\bibitem{99:GottesmanB}
D. Gottesman and I.~L. Chuang, Nature {\bf 402},  390  (1999),
  quant-ph/9908010.

\bibitem{97:KitaevA}
A.~Y. Kitaev,  in {\em Quantum Communication, Computing and Measurement (Proc.
  3rd Int. Conf. of Quantum Communication and Measurement)} (Plenum Press, New
  York, 1997), pp.\ 181--188.

\bibitem{98:Knill}
E. Knill, R. Laflamme, and W.~H. Zurek, Science {\bf 279},  342  (1998).

\bibitem{99:SteaneB}
A.~M. Steane, Nature {\bf 399},  124  (1999), quant-ph/9809054.

\bibitem{97:SteaneC}
A.~M. Steane, Fortschritte der Physik (Prog. Phys.) {\bf 46},  443  (1997),
  quant-ph/9708021.

\bibitem{02:SteaneA}
A.~M. Steane, submitted for publication  (2002), quant-ph/0202036.

\bibitem{98:Aharonov}
D. Aharonov and M. Ben-Or,  in {\em Proc. 29th Ann. ACM Symp. on Theory of
  Computing} (ACM, New York, 1998), p.\ 176, quant-ph/9906129,
  quant-ph/9611025.

\bibitem{97:SteaneA}
A.~M. Steane, Phys. Rev. Lett. {\bf 78},  2252  (1997), quant-ph/9608026.

\bibitem{97:KitaevB}
A.~Y. Kitaev,   (1997), quant-ph/9707021.

\bibitem{Th:Gottesman}
D. Gottesman, Ph.{D}.~thesis, California Institute of Technology, 1997.

\bibitem{02:SteaneB}
A.~M. Steane, Quant. Inf. and Comp. {\bf 2},  297  (2002), quant-ph/0203047.

\bibitem{01:SteaneA}
A.~M. Steane,  in {\em Decoherence and its Implications in Quantum Computation
  and Information Transfer} (IOS Press, Amsterdam, 2001), pp.\ 284--298,
  quant-ph/0304016.

\bibitem{97:Knill}
E. Knill and R. Laflamme, Phys. Rev. A {\bf 55},  900  (1997),
  quant-ph/9604034.

\bibitem{00:KnillB}
E. Knill, R. Laflamme, and L. Viola, Phys. Rev. Lett. {\bf 84},  2525  (2000),
  quant-ph/9908066.

\bibitem{01:Alicki}
R. Alicki, M. Horodecki, P. Horodecki, and R. Horodecki, PhysRevA {\bf 65},
  062101  (2002), quant-ph/0105115.

\bibitem{99:Zalka}
C. Zalka,   (1999), quant-ph/9612028 v2.

\bibitem{Bk:Press}
W.~H. Press, S.~A. Teukolsky, W.~T. Vetterling, and B.~P. Flannery, {\em
  Numerical Recipes in {C}, the art of scientific computing} (Cambridge
  University Press, England, 1994).

\bibitem{97:Preskill}
J. Preskill,  in {\em Introduction to Quantum Computation}, edited by H.-K. Lo,
  S. Popescu, and T. Spiller (World Scientific, Singapore, 1998), pp.\
  213--269, quant-ph/9712048.

\bibitem{Bk:Macwilliams}
F.~J. Mac{W}illiams and N.~J.~A. Sloane, {\em The Theory of Error-Correcting
  Codes} (North-Holland, Amsterdam, 1977).

\bibitem{96:SteaneC}
A.~M. Steane, Phys. Rev. A {\bf 54},  4741  (1996).

\bibitem{98:Calderbank}
A.~R. Calderbank, E.~M. Rains, P.~W. Shor, and N.~J.~A. Sloane, IEEE
  Transactions on Information Theory {\bf 44},  1369  (1998).

\end{thebibliography}

\end{document}